\documentclass[aps,pra,twocolumn,twoside]{revtex4}
\usepackage[utf8]{inputenc}
\usepackage{graphicx}
\usepackage[pdftex,hypertexnames=false,pdfpagemode=UseThumbs,linkbordercolor={1 1 1}, citebordercolor={1 1 1},pdfstartview=FitH]{hyperref}
\usepackage{amsmath}
\usepackage{bbold}
\DeclareMathAlphabet{\mathpzc}{OT1}{pzc}{m}{it}
\renewcommand{\vec}[1]{\boldsymbol{#1}}
\begin{document}
\title{Spintronics with graphene quantum dots}
\author{Matthias Droth}
\author{Guido Burkard}
\affiliation{Department of Physics, University of Konstanz, 78457 Konstanz, Germany}
\begin{abstract}
Thanks to its intrinsic ability to preserve spin coherence, graphene is a prime material for spintronics. In this review article, we summarize recent achievements related to spintronics in graphene quantum dots and motivate this field from a spintronics and a materials science point of view. We focus on theory but also discuss recent experiments. The main sources of spin decoherence are interactions with lattice excitations and the hyperfine interaction with present nuclear spins. We explain effective spin-phonon coupling in detail and present a generic power law for the spin relaxation time $T_1$ as a function of the magnetic field. For specific cases, we discuss spin relaxation in detail. The Heisenberg exchange interaction is paramount for coherent spin qubit operation and addressed in the context of magnetism in graphene nanoflakes. Nuclear spins in the host and surrounding material can be considered by several means and the influence of $^{13}$C nuclei has been studied in detail. Impressive advances in general spintronics and the fabrication of graphene devices are likely to spark significant advances in spintronics with graphene quantum dots in the near future.
\end{abstract}
\maketitle
\section{Introduction}
Spintronics and graphene are exciting, vigorous, and rather new areas of research with a rapid pace of new discoveries. Books and a multitude of review articles are available for those that want to learn about one of these fields [\onlinecite{Wolf2001,Awschalom,Zutic2004,CastroNeto2009,Katsnelson,Guclu,Awschalom2013,Sinova2012,Jansen2012}]. The intersection of both provides fertile soil for an abundance of fascinating physics and its fruits have been partially reaped by previous reviews [\onlinecite{Recher2010,Pesin2012,Han2014}]. Here, we put the cherry on the cake and review specifically spintronics in graphene quantum dots.

While classical electronics relies on the charge for information processing, spin-based electronics or \emph{spintronics} is the paradigm of an advanced technology where the spin degree of freedom complements or even replaces charge as the carrier of information. The expected benefits of spintronics encompass non-volatile data storage, faster and more energy efficient data processing, increased data density, and many more. In order to fully exploit the potential, efficient generation, transport, transfer, manipulation, and detection of spin polarization is required. All these requirements are closely connected to material properties. Therefore, progress in spintronics seems intrinsically connected to the quest for new materials with appropriate characteristics [\onlinecite{Wolf2001,Jansen2012,Han2014}].

\addtocounter{figure}{-1}
\begin{figure}[t]\centering\includegraphics[width=\linewidth]{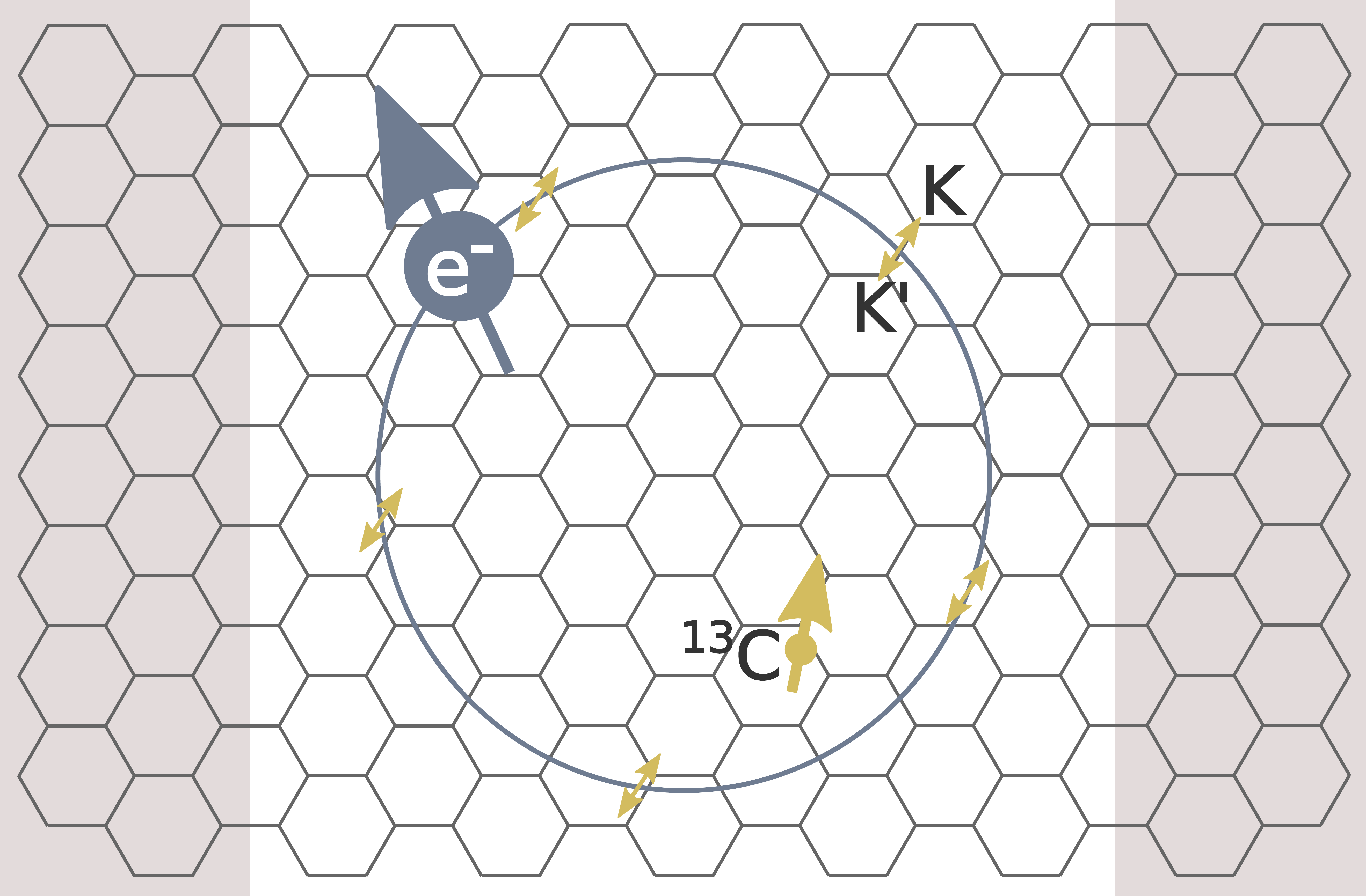}
\caption{An electron spin in a graphene quantum dot interacts with the electron momentum via the spin-orbit coupling and with the 
sparse $^{13}$C nuclear spin via the hyperfine interaction.}
\end{figure}

Graphene is a novel material that has undeniably stirred up the solid state community since its isolation more than a decade ago [\onlinecite{Jang2002Pat,Novoselov2004,Geim2007,Savage2012}]. It is a monatomically thin, quasi two-dimensional layer of carbon atoms arranged in a honeycomb lattice with two sublattices, A and B. Due to its atomic structure, it has the lowest surface mass density of all solids [\onlinecite{Brodie1859}]. At the same time, its in-plane mechanical strength is second to no other material [\onlinecite{Lee2008}]. The hexagonal structure is the consequence of sp$_2$-hybridized carbon atoms. The remaining p-orbitals form the $\pi$-bands responsible for transport. Electrons move quasi-relativistically with the Fermi velocity $v_{\rm F}{\approx}c/300$, where $c$ is the vacuum speed of light, and with high mobilities [\onlinecite{Wallace1947,DiVincenzo1984,Semenoff1984,Novoselov2005,Lin2010}]. Spin-orbit couplings are small and nuclear spins are sparse, making it a potentially good host material for spin [\onlinecite{Trauzettel2007,Kane2005,Min2006,Gmitra2009,Konschuh2010,Audi2003}]. In addition to the real electron spin, the occupation of sublattices A and B can be described by the sublattice pseudospin. Moreover, in momentum space, the occupation of the two inequivalent Dirac points --- where valence and conduction bands touch --- may be denoted by the valley isospin [\onlinecite{CastroNeto2009,Pesin2012}]. Electron spin currents can be injected optically in graphene with Rashba-type spin-orbit interaction [\onlinecite{Nair2008,Rioux2014,Inglot2014}]. This allows to leave away ferromagnetic contacts, thus eliminating a possible source of spin scattering. Magnetic behavior has been observed at the edges of graphene [\onlinecite{Yazyev2007,Yazyev2010,Tao2011}]. These and further properties make graphene a very promising material for spintronics.

Quantum dots (QDs), in graphene as well as in other materials, are quasi zero-dimensional regions to which charge carriers can be confined [\onlinecite{Guclu,Recher2010,Tarucha1996,Kouwenhoven2001}]. With electrostatic gate electrodes that can be used to adjust the confinement potential and with electric contacts for transport, QDs allow in principle for full control over the individual electron. In particular, QDs provide a controllable playground to investigate the behavior of spins as well as sources of decoherence and methods to increase coherence times [\onlinecite{Awschalom2013,Hanson2007,Petta2005}]. In general, there are nuclear spins as well as electron spins in a solid. Unless explicitly stated otherwise, we always refer to electron spin.

In this review, we focus on what we believe to be the most immediate questions related to spintronics with localized electron spins in graphene. However, we point out that there are a several related systems that make use of carbon nanotubes or transition metal dichalcogenides (TMDCs) and where the electron spin degree of freedom might be substituted or complemented by a pseudospin or valley isospin [\onlinecite{Pesin2012,Winkler2015,Palyi2012,Wang2014,Palyi2011,Rohling2014,Xu2014,Kormanyos2014}]. The review is organized as follows. In the next section, we give an overview over spintronics in quantum dots and motivate the use of graphene. In Sec.~\ref{T1sec}, we discuss various graphene quantum dots and their associated spin relaxation times. Closely related to the concept of QDs are localized states due to defects in graphene nanoflakes, which we treat in Sec.~\ref{GNF}. In Sec.~\ref{NSpins}, we address the influence of nuclear spins on spintronics in graphene. Recent experimental achievements and the current status are the topic of Sec.~\ref{Exp}. Finally, we give a summary and perspective in Sec.~\ref{CandO}.

\section{Spintronics}
Quantum effects become increasingly relevant as components of information processing devices shrink to the few-nanometer scale. This poses a challenge but also offers the chance to exploit quantum mechanics for data processing [\onlinecite{Feynman1960,Moore1965,Thompson2006}]. The spin is the canonical example of a quantum mechanical degree of freedom with two eigenstates. In solids, it is possible to use spin-orbit coupling and many other spin dependent effects to manipulate the spin with electric fields and hence much faster than with external magnetic fields. In order to prepare, manipulate, and detect spin in an active and controlled way, spintronic devices may take advantage of any such effects [\onlinecite{Dyakonov1971,Datta1990,Kato2004,Bhowmik2014}].

In contrast to electronics where carrier mobilities and lifetimes matter, spin mobilities and coherence times are relevant parameters for spintronics. The coherence of electron spins is mainly determined by the spin-orbit interaction (SOI), the abundance of nuclear spins in the host material, and magnetic behavior. These characteristics depend on the material and interlink spintronics with material science. Due to their lower magnetic moment, nuclear spins couple weakly to the environment and exhibit typical coherence times orders of magnitude longer than for electron spins [\onlinecite{Wolf2001,Morton2008}].

Spintronics promises new applications and substantially better performance than standard micro- and nanoelectronics by exploiting the spin degree of freedom. Hard drive read heads are a multibillion dollar industry and magnetoresistive random-access memory (MRAM) is still waiting in the wings. Both technologies are based on the giant magnetoresistant effect, whose discovery in the late 1980s may be considered the birth of spintronics [\onlinecite{Baibich1988,Binasch1989,Park2011}]. Beyond classical computing, spintronics has the potential for key components of quantum computing [\onlinecite{Loss1998,NielsenChuang}]. Quantum computing can be viewed as a coherent superposition of classical computations running in parallel, with algorithms like the one by Shor exploiting this [\onlinecite{Shor1994}]. Obviously, such algorithms require appropriate quantum hardware [\onlinecite{DiVincenzo1998}].
\subsection{Spins in quantum dots}
As proposed by Loss and DiVincenzo, spin states of coupled QDs can be used to implement a universal set of one- and two-qubit gates (\emph{qubit} is short for quantum bit). Spin qubit gates can be realized with magnetic fields and, using spin-dependent effects, also with electric fields, or a combination of both [\onlinecite{Loss1998,DiVincenzo1998}]. Such a solid state approach to quantum computing benefits from the scalability known from standard electronics and from long coherence times [\onlinecite{Awschalom2013,Tyryshkin2012,Shulman2012,Takakura2014}].

The dot regions to which an integer number of electrons can be confined are typically on the length scale of 10-100 nanometers. The eigenstate spectrum is discrete and quantum dots are often referred to as \emph{artificial atoms}. Coulomb repulsion and the exchange coupling also play a role [\onlinecite{Tarucha1996,Kouwenhoven2001,Hanson2007}].

Also for spins that are confined to the QD, there remains a coupling to the environment. At finite temperatures, spins and phonons in the host material couple via the admixture mechanism, which involves spin-orbit interaction and electron-phonon coupling, or directly, via spin-orbit interaction only [\onlinecite{Khaetskii2001,Rudner2010}]. Moreover, the hyperfine interaction mediates a coupling between the electron spin and nuclear spins in its vicinity [\onlinecite{Hanson2007,Coish2009}]. These mechanisms lead to a finite spin lifetime, an important figure of merit for applications like the Loss-DiVincenzo quantum computer where it should exceed the clock time by orders of magnitude [\onlinecite{DiVincenzo1998,Barends2014}]. The theoretical prediction and explanation of spin lifetimes (in graphene QDs) is part of this review.

\subsection{Why use graphene for spintronics?}\label{why}
Its many outstanding properties make graphene, a monatomically thin layer of carbon atoms, an interesting material for a variety of applications, including spintronics. The low spin-orbit coupling and a small abundance of nuclear spins are beneficial for long spin coherence times [\onlinecite{Trauzettel2007}]. In Table \ref{t1}, we compare the main sources of decoherence for different materials.

Electron mobilities in graphene are typically high and allow for spin transport over micrometer length scales [\onlinecite{Lin2010,Guimaraes2012,Han2012,Drogeler2014,Kamalakar2015}]. While graphene is naturally non-magnetic, it can exhibit magnetic behavior under certain conditions [\onlinecite{Yazyev2007,Yazyev2010,Tao2011,Droth2015}]. 
More recently, there has been an effort to stack graphene and other two-dimensional materials like hexagonal boron nitride (hBN) and TMDCs on top of each other, held together by van der Waals forces [\onlinecite{Novoselov2012,Geim2013,Hunt2013,Cui2015}]. It is envisioned that by combining different materials in such van der Waals heterostructures, one may tailor physical properties. The concept is akin to bandgap engineering in semiconductor alloys and might once again highlight the connection between spintronics and materials science.

The hexagonal structure of graphene comes from the sp$_2$-hybrid formed by three of the four outer-shell electrons in carbon. The resulting $\sigma$-bands are energetically far away from the charge neutrality point and play no role in transport. The remaining p$_z$ orbitals form the $\pi$-bands responsible for transport. In the low-energy approximation, they obey the quasi-relativistic Dirac Hamiltonian [\onlinecite{CastroNeto2009}]
\begin{equation}
\mathcal{H}_{\rm D}=\hbar v_{\rm F}(\tau\sigma_xq_x+\sigma_yq_y)+\Delta\sigma_z+V(x,y)\,,\label{Dirac}
\end{equation}
where $\tau$ labels the Dirac valleys $K$ ($\tau{=}{+}1$) and $K'$ ($\tau{=}{-}1$) in reciprocal space, $\vec{\sigma}$ are Pauli matrices in standard representation that describe the sublattice, and $\vec{q}$ is the wave vector measured w.r.t.~$K$ or $K'$, respectively. The sublattice splitting $\Delta$ vanishes for ideal graphene but may occur due to a substrate [\onlinecite{Hunt2013}]. An electrostatic potential may be considered via the term $V(x,y)$. The eigenspectrum of Eq.~(\ref{Dirac}) with $V{=}0$ is given by
\begin{equation}
E=\pm\sqrt{(\hbar v_{\rm F})^2(q_x^2+q_y^2)+\Delta^2}\label{spectrum}
\end{equation}
and has no gap if $q{=}0$ is possible and $\Delta{=}0$. In that case, it is not possible to confine electrons electrostatically because of Klein tunneling [\onlinecite{Klein1929,Katsnelson2006}]. In Sec.~\ref{T1sec}, we will discuss special cases where the energy spectrum becomes gapped, thus enabling electrostatic confinement.

\begin{table}[t]
\centering
\caption{Sources of decoherence in different materials. If two spin-orbit splittings are listed, the upper (lower) number corresponds to the conduction (valence) band. Data on the different isotopes is taken from Ref.~[\onlinecite{Audi2003}].\\
$\dag$ Data for a (4,4) armchair carbon nanotube. The spin-orbit coupling scales inversely with the nanotube radius [\onlinecite{HuertasHernando2006}].\\
$\ddag$ The isotopes $^{92}$Mo, $^{94}$Mo, $^{96}$Mo, $^{98}$Mo, and $^{100}$Mo have nuclear spin 0 and a combined natural abundance of 75\%.}
\begin{tabular}{lll}
\hline
Material & Spin-orbit splitting & Nuclear spin \\
\hline
Graphene&$24\,\text{$\mu$eV}$ [\onlinecite{Gmitra2009,Abdelouahed2010}]&$0$ (99\% $^{12}$C)\\
&&$\frac{1}{2}$ (1\% $^{13}$C)\\
Bilayer graphene&$24\,\text{$\mu$eV}$ [\onlinecite{Konschuh2012}]&\\
Carbon nanotube$^{\dag}$&$1.6\,\text{meV}$ [\onlinecite{Han2014}]&\\
\hline
Silicene&$1.6\,\text{meV}$ [\onlinecite{Liu2011}]&$0$ (95\% $^{28,30}$Si)\\
&&$\frac{1}{2}$ (5\% $^{29}$Si)\\
Silicon (3D bulk)&$44\,\text{meV}$ [\onlinecite{Winkler}]&\\
\hline
2D hexagonal BN&$15\,\text{$\mu$eV}$ [\onlinecite{Han2014}]&$\frac{3}{2}$ (80\% $^{11}$B)\\
&$30\,\text{$\mu$eV}$ [\onlinecite{Han2014}]&$3$ (20\% $^{10}$B)\\
&&$1$ ($>$99\% $^{14}$N)\\
&&$\frac{1}{2}$ ($<$1\% $^{15}$N)\\
\hline
2D MoS$_2$&$3\,\text{meV}$ [\onlinecite{Kosmider2013}]&$0$ (75\% $^{\ddag}$Mo)\\
&$147\,\text{meV}$ [\onlinecite{Kosmider2013}]&$\frac{5}{2}$ (25\% $^{95,97}$Mo)\\
&&$0$ (99\% $^{32,34,36}$S)\\
&&$\frac{3}{2}$ (1\% $^{33}$S)\\
\hline
GaAs (3D bulk)&$340\,\text{meV}$ [\onlinecite{Winkler}]&$\frac{3}{2}$ (100\% $^{69,71}$Ga)\\
&&$\frac{3}{2}$ (100\% $^{75}$As)\\
\hline
\end{tabular}
\label{t1}
\end{table}

Due to its low nuclear charge, carbon has a relatively low atomic spin-orbit interaction and this carries through to the spin-orbit interactions for band electrons in all carbon based materials. Yet for flat graphene, the coupling is particularly weak because the $\pi$- and $\sigma$-bands are orthogonal [\onlinecite{Kane2005,Min2006,Gmitra2009,Konschuh2010}]. The spin-orbit coupling in graphene is given by
\begin{equation}
\mathcal{H}_{\rm SOI}=\lambda_{\rm I}\tau\sigma_zs_z+\lambda_{\rm R}(\tau\sigma_xs_y-\sigma_ys_x)\,,\label{SOI}
\end{equation}
where the first term denotes the intrinsic (or Dresselhaus-type) effect $\mathcal{H}_{\rm I}$ and the second term stands for the extrinsic (or Rashba-type) effect $\mathcal{H}_{\rm R}$, which may be induced by a substrate or an external electric field. The coupling strengths are not completely settled, but for concreteness we list $\lambda_{\rm I}{=}12\,\text{$\mu$eV}$ and $\lambda_{\rm R}{=}5\,\text{$\mu$eV}\times E\,[\text{V/nm}]$ (see Ref.~[\onlinecite{Gmitra2009}] and also Table \ref{t1}).

The two-dimensional surface states of three-dimensional topological insulators (TIs) are similar to graphene in that their low-energy excitations are also described by a Dirac Hamiltonian. While spin-orbit coupling is flimsy in graphene, it is strong in TIs. Future spintronics devices may thus rely on graphene for spin transport, combined with TIs for spin manipulation [\onlinecite{Pesin2012}]. However, spin relaxation times measured in transport experiments in graphene are significantly shorter than expected and the underlying mechanisms remain somewhat elusive [\onlinecite{Han2014,Guimaraes2012,Tombros2007}]. Nevertheless, the past few years have brought many new insights and experimental spin lifetimes have increased with improvements in device fabrication [\onlinecite{Kamalakar2015,Boross2013,Diez2012,Balakrishnan2013,Fedorov2013}].

Since protons are 1836 times more massive than electrons, their magnetic moments are smaller than the Bohr magneton by that factor. This inhibits the exchange of angular momentum between electron spins and nuclear spins if a magnetic field is involved as the resulting Zeeman splitting scales linear with the magnetic moment. Still, nuclear spins do play an important role for spin coherence [\onlinecite{Coish2009}]. The natural abundance of carbon isotopes is dominated by 99\% $^{12}$C, which has no nuclear spin. About 1\% of carbon atoms are $^{13}$C isotopes and have nuclear spin $\frac{1}{2}$. With the atomic distance of $1.42\,\text{\AA}$ in graphene, one thus expects one atom with nuclear spin $\frac{1}{2}$ in a square area of $(16\,\text{\AA})^2$. While such small QDs can be realized by electroburning [\onlinecite{Barreiro2012}], typical QD dimensions are rather $(16\,\text{nm})^2$ or $(160\,\text{nm})^2$, according to $100$ to $10\,000$ atoms with nuclear spin [\onlinecite{Liu2010,Allen2012,Goossens2012,Varlet2014,Tan2015}]. For silicon, whose natural abundance of isotopes is dominated by isotopes without nuclear spin (92\% $^{28}$Si, 3\% $^{30}$Si), electron spin coherence times exceeding seconds have been achieved after isotopic purification [\onlinecite{Tyryshkin2012}]. Given that natural carbon has even fewer nuclear spins than natural silicon (see Table \ref{t1}), this technique would be most suitable for graphene [\onlinecite{Balasubramanian2009}].

In addition to the real spin, electrons in graphene have other binary degrees of freedom, namely the sublattice spin that refers to sublattices A and B (usually referred to as \emph{pseudospin}) and the valley spin that refers to the Dirac points $K$ and $K'$ (technically also a \emph{pseudo}spin but usually called \emph{isospin}). Moreover, the two layers in bilayer graphene can be attributed a layer spin. The valley spin also occurs in TMDCs and valleytronics --- the electric initialization, manipulation, and detection of valley spin --- is an active area of research [\onlinecite{Pesin2012,Rohling2014,Xu2014,Rohling2012}]. Now, we turn to graphene quantum dots and the lifetimes of electron spins confined to those QDs.

\section{Graphene quantum dots and spin relaxation}\label{T1sec}
There are several possibilities to localize electrons to a well-defined area within a graphene sheet. One possibility is to lithographically cut the desired shape of a graphene island into the 2D bulk, to which it remains connected by tiny nanoconstrictions. In this case, boundary effects need to be under control to understand and manipulate the behavior of electrons such QDs [\onlinecite{Ponomarenko2008,Schnez2010}]. Other approaches involve inhomogeneous magnetic fields or rely on localization due to disorder [\onlinecite{DeMartino2007,Pereira2006}].

\begin{figure}[t]\centering\includegraphics[width=0.8\linewidth]{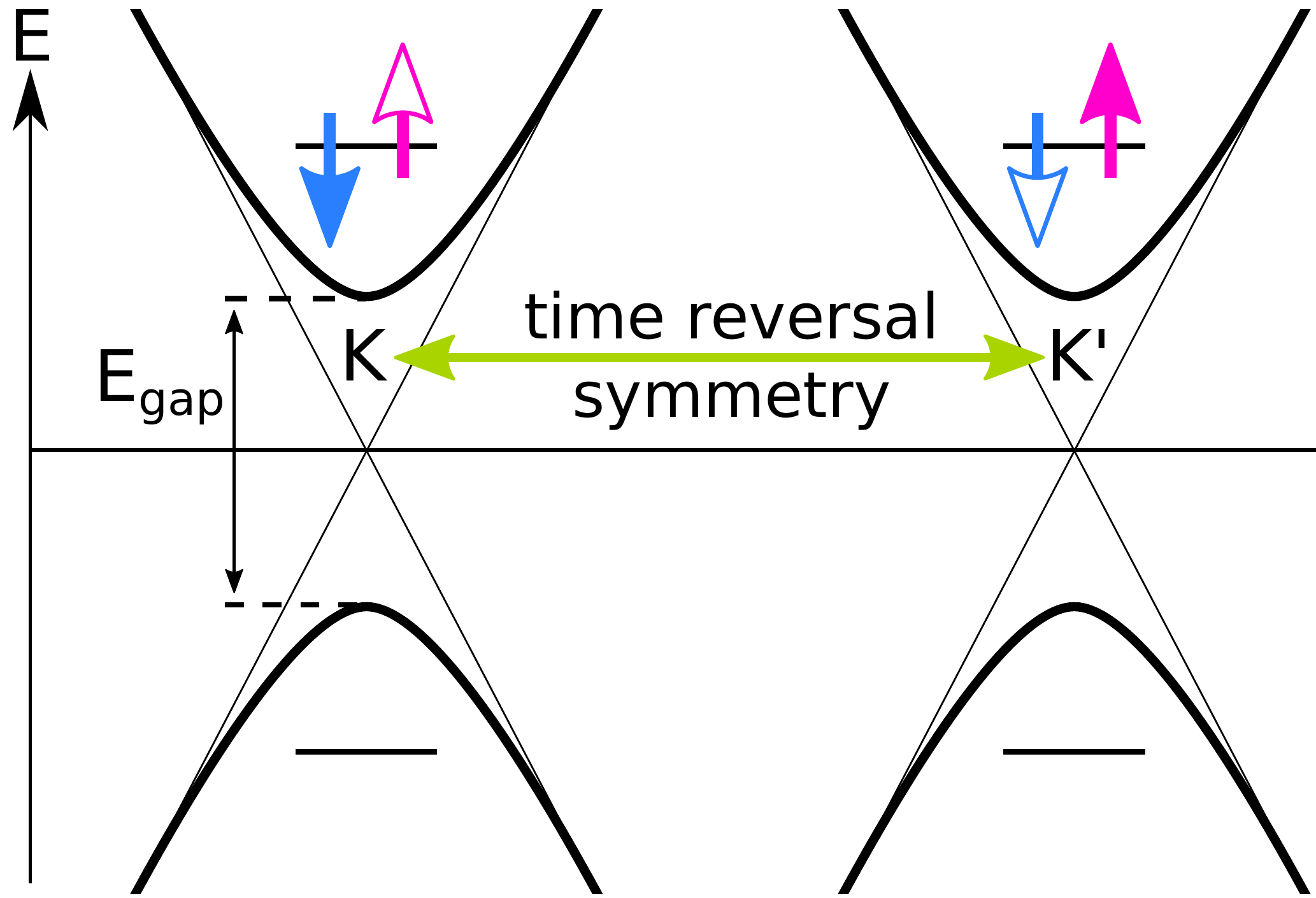}\caption{The electronic spectrum poses two challenges for spintronics with graphene quantum dots. (i) Pristine graphene has a linear, gapless dispersion (thin lines) that allows charge carriers to override electrostatic barriers. A finite energy gap is necessary to avoid this phenomenon known as the Klein paradox [\onlinecite{Klein1929,Katsnelson2006}]. (ii) The orbital states at valleys $K$ and $K'$ are connected by time reversal symmetry and hence twofold degenerate. Together with spin degeneracy, this valley degeneracy makes for a fourfold degenerate system rather than a two-level qubit, with an accordingly more complicated exchange mechanism [\onlinecite{Rohling2012}]. This can be avoided by lifting the valley degeneracy. If $K$ and $K'$ are degenerate due to time reversal symmetry, there will be two sets of Kramers qubits (solid and empty arrows, respectively), i.e., qubits that are connected by time reversal symmetry. Kramers spin qubits have long lifetimes due to the Van Vleck cancellation [\onlinecite{Khaetskii2001,VanVleck1940}].}\label{pic1}\end{figure}

In order to benefit from the expertise with semiconductor QDs and to facilitate fast switching times, electrostatic confinement of electrons in gate-tunable QDs is desirable. For electron spins, the Dirac Hamiltonian implies two challenges, depicted in Fig.~\ref{pic1}: first (i), the spectrum, given by Eq.~(\ref{spectrum}), must be gapped in order to avoid Klein tunneling, i.e., transmission of conduction electrons through the electrostatic barriers via the valence band states [\onlinecite{Katsnelson2006}]. (ii) Due to spin and valley degeneracies, the orbital states of graphene electrons are fourfold degenerate, i.e., they form four-level quantum systems rather than two-level qubits. Therefore, the time-reversal symmetry between Dirac points $K$ and $K'$ needs to be lifted unless one accepts a complicated exchange interaction for the four-level system, which is also possible [\onlinecite{Rohling2012}]. Both challenges are met by graphene nanoribbons with a certain type of boundary condition [\onlinecite{Recher2010,Trauzettel2007}] as well as by circular quantum dots in mono- or bilayer graphene and with finite magnetic field [\onlinecite{Recher2009}].

In the following subsection, we introduce general concepts of spin relaxation via the spin-orbit interaction mediated coupling to lattice vibrations. After that, we discuss details of the mentioned graphene QDs with lifted valley degeneracies.

%

\subsection{Spin-orbit mediated spin relaxation}
If spin decoherence is dominated by interactions with phonons, nuclear spins can be neglected and the system can be described by the Hamiltonian
\begin{equation}
\mathcal{H}=\mathcal{H}_{\rm elec}+\mathcal{H}_{\rm phon}+\mathcal{H}_{\rm soi}+\mathcal{H}_{\rm epc}\,,
\end{equation}
where $\mathcal{H}_{\rm elec}$ and $\mathcal{H}_{\rm phon}$ model the unperturbed electronic and vibrational systems, respectively [\onlinecite{footnote}]. The spin-orbit interaction $\mathcal{H}_{\rm soi}{=}\mathcal{H}_{\rm r}{+}\mathcal{H}_{\rm i}$ is the sum of extrinsic (or Rashba-type) coupling and intrinsic (or Dresselhaus-type) coupling. Together with the electron-phonon coupling $\mathcal{H}_{\rm epc}$, it mediates the coupling between spins and the lattice phonons. We denote the orbital eigenstates of $\mathcal{H}_{\rm elec}$ as $|n\rangle$. For graphene, $\mathcal{H}_{\rm elec}$ is usually given by some form of Eq.~(\ref{Dirac}). The pure vibrational modes are described by
\begin{equation}
\mathcal{H}_{\rm phon}=\sum_{\alpha,\vec{q}}\hbar\omega_{\alpha,\vec{q}}\left(m_{\alpha,\vec{q}}+\frac{1}{2}\right)\,,
\end{equation}
where the summation runs over all acoustic phonon branches $\alpha$ and wave vectors $\vec{q}$. The angular frequency $\omega_{\alpha,\vec{q}}$ of a harmonic vibrational mode is implicitly determined by $\alpha$ and $\vec{q}$, and $m_{\alpha,\vec{q}}$ is the occupation number of this mode. The according eigenstates are determined by these occupation numbers, $|m_{\alpha,\vec{q}}\rangle$.

\begin{figure}[t]\centering\includegraphics[width=1\linewidth]{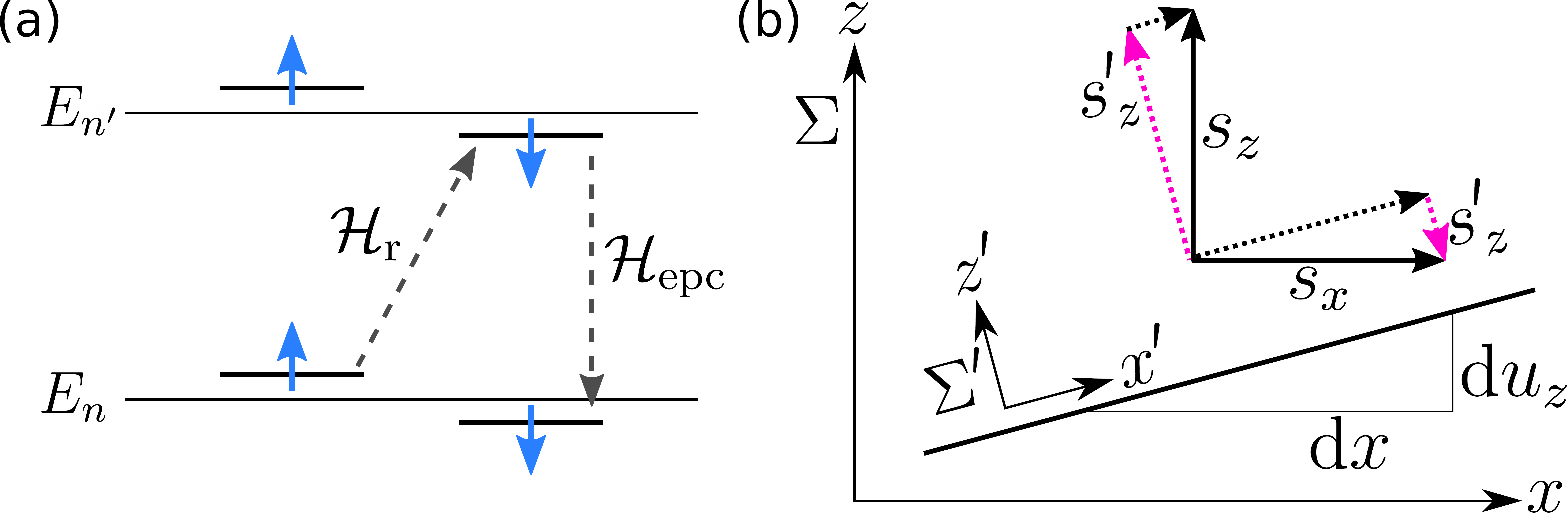}\caption{(a) The admixture mechanism allows an electron in orbital state $|n\rangle$ to flip its spin via a virtual transition to a different orbital state $|n'\rangle$. Both electron-phonon coupling $\mathcal{H}_{\rm epc}$ and Rashba-type spin-orbit interaction $\mathcal{H}_{\rm r}$ couple the different orbits but only $\mathcal{H}_{\rm r}$ leads to a spin flip, thus effectively enabling a spin flip within the orbital state $|n\rangle$. (b) Due to out-of-plane phonons, the local coordinate frame of the graphene lattice, $\Sigma'$, that describes pseudospin $\vec{\sigma}'$ may differ from the laboratory frame $\Sigma$ that fixes the quantization axis of the electron spin $\vec{s}$ in an external magnetic field. In the local lattice frame, $s_z'$ may depend on all components of $\vec{s}$, as indicated by the red arrows. This changes the intrinsic spin-orbit coupling to the form of Eq.~(\ref{directSPC}), where the primes have been dropped.}\label{pic2}\end{figure}

Phonons are bosons with zero spin and hence cannot couple to electron spin directly but via the \emph{admixture mechanism} which combines Rashba-type spin-orbit interaction $\mathcal{H}_{\rm r}$ and electron-phonon coupling $\mathcal{H}_{\rm epc}$ [\onlinecite{Khaetskii2001}]. Let $|n,s\rangle^{(0)}$ be the product state of an electron in the orbital state $|n\rangle$ and with spin $|s\rangle$. In lowest order, $\mathcal{H}_{\rm r}$ admixes product states with opposite spin,
\begin{equation}
|n\uparrow\rangle{=}|n\uparrow\rangle^{(0)}{+}\sum_{n'\neq n}|n'\downarrow\rangle^{(0)}\frac{^{(0)}\langle n'\downarrow|\mathcal{H}_{\rm r}|n\uparrow\rangle^{(0)}}{E_n{-}E_{n'}{+}g\mu_BB}\,,\label{1stOrder}
\end{equation}
where $E_n$ is the orbital energy and $\pm g\mu_BB$ is the Zeeman energy with the $g$-factor ($g{=}2$ for graphene [\onlinecite{Zhang2006}]), the Bohr magneton $\mu_{\rm B}$, and a magnetic field $B$. An according expression for $|n\downarrow\rangle$ leads to finite spin-flipping matrix elements $\langle n\downarrow|\mathcal{H}_{\rm epc}|n\uparrow\rangle{=}$
\begin{equation}
\sum_{n'\neq n}\left(\frac{(\mathcal{H}_{\rm epc})_{nn'}(\mathcal{H}_{\rm r})_{n'n}^{\downarrow\uparrow}}{E_n{-}E_{n'}{+}g\mu_BB}{+}\frac{(\mathcal{H}_{\rm epc})_{n'n}(\mathcal{H}_{\rm r})_{nn'}^{\downarrow\uparrow}}{E_n{-}E_{n'}{-}g\mu_BB}\right)\,,\label{admix}
\end{equation}
where we denote the numerator in Eq.~(\ref{1stOrder}) as $(\mathcal{H}_{\rm r})_{n'n}^{\downarrow\uparrow}$ and the spin-conserving transitions of $(\mathcal{H}_{\rm epc})_{n'n}$ accordingly. The working principle of the admixture mechanism is illustrated in Fig.~\ref{pic2} (a). With $\rho(\hbar\omega_{\alpha,\vec{q}})$ as the phonon density of states, the relaxation rate can now be calculated via Fermi's golden rule,
\begin{equation}
T_1^{-1}{=}\frac{2\pi}{\hbar}\sum_{\alpha,\vec{q}}|\langle n\downarrow,m_{\alpha,\vec{q}}{+}1|\mathcal{H}_{\rm epc}|n\uparrow,m_{\alpha,\vec{q}}\rangle|^2\rho(\hbar\omega_{\alpha,\vec{q}})\,.\label{Fermi}
\end{equation}

\begin{table}[t]
\centering
\caption{Overview of the parameters used for the generic model of $T_1^{-1}(B)$ in Eq.~(\ref{T1estimate}).}
\begin{tabular}{lll}
\hline
Description & Parameter & Value\\
\hline
Dipole approximation&$\alpha$&0 (zeroth order)\\
&&1 (first order)\\
\hline
Kramers pair&$\beta$&0 (no)\\
&&1 (yes)\\
\hline
System dimension&$D$&1\\
&&2\\
&&3\\
\hline
Phonon dispersion&$\eta$&1 (linear)\\
&&2 (quadratic)\\
&&... (higher)\\
\hline
\end{tabular}
\label{t2}
\end{table}

Due to energy conservation, the phonon energy must match the Zeeman splitting, $\hbar\omega_{\alpha,\vec{q}}{=}g\mu_BB$. For typical laboratory magnetic fields, $B{\lesssim}30\,\text{T}$, this implies long-wavelength acoustic phonons at the center of the Brillouin zone that can be described by the continuum model. In a very general picture, the atomic displacement in the presence of an acoustic phonon behaves as $\vec{u}_{\alpha,\vec{q}}(\vec{r}){\propto}\omega_{\alpha,\vec{q}}^{-\frac{1}{2}}e^{i(\vec{q}\cdot\vec{r}-\omega_{\alpha,\vec{q}}t)}$, where the prefactor occurs in the normal coordinate for any harmonic oscillator. A constant acoustic displacement $\vec{u}$ corresponds to a mere translation of the lattice and hence does not give rise to electron-phonon coupling. Usually, $\mathcal{H}_{\rm epc}$ involves derivatives $\partial_iu_j$, see, e.g., Eqs.~(\ref{EPC}) and (\ref{directSPC}), or Ref.~[\onlinecite{Khaetskii2001}].

The matrix element given by Eq.~(\ref{admix}) scales with $q^{1+\alpha}B^{\beta-\frac{1}{2}}$, where one power in $q$ comes from these spatial derivatives. In the long wavelength limit, the factor $e^{i\vec{q}\cdot\vec{r}}$ can be replaced with the dipole approximation $e^{i\vec{q}\cdot\vec{r}}{\approx}1{+}i\vec{q}{\cdot}\vec{r}$. Depending on the symmetry of the electronic states, the constant term may ($\alpha{=}0$) or may not ($\alpha{=}1$) give a contribution to the matrix element. If the two states of the spin qubit are related by time reversal symmetry in the zero field limit, i.e., if they form a Kramers pair (Fig.~\ref{pic1}), the matrix element will vanish unless the Zeeman energy in the denominators is taken into account [\onlinecite{Khaetskii2001,VanVleck1940}]. Therefore, $\beta{=}1$ ($\beta{=}0$) if the spin states form a (no) Kramers pair. Finally, the factor $B^{-\frac{1}{2}}$ comes from the normal coordinate of the acoustic phonon.

The density of states in Eq.~(\ref{Fermi}) scales with $A\frac{\text{d}q}{\text{d}\omega}$, where $A{\propto}q^{D-1}$ is the content of the iso-energy hyperplane in $D$-dimensional reciprocal space ($D{\in}\{1,2,3\}$). Assuming a phonon dispersion $\omega{\propto}q^{\eta}$ ($\eta{\in}\mathbb{N}$), the derivative scales as $\omega^{\frac{1-\eta}{\eta}}$ and hence $\rho{\propto}B^{\frac{D}{\eta}-1}$, where we have used $\omega{\propto}B$. Using $q{\propto}B^{\frac{1}{\eta}}$ for the matrix element, too, we find a general estimate for Eq.~(\ref{Fermi}),
\begin{equation}
T_1^{-1}{\propto}B^{2\frac{1+\alpha}{\eta}}B^{2(\beta-1)}B^{\frac{D}{\eta}}\,.\label{T1estimate}
\end{equation}
An overview of the parameters in this formula is provided in Table \ref{t2} and specific examples are shown in Table \ref{t3}. Unless direct restoring forces are absent (as, e.g., for out-of-plane modes in free graphene [\onlinecite{Falkovsky2008,Droth2011}]) acoustic modes disperse linearly ($\eta{=}1$) and the above relation simplifies to $T_1^{-1}{\propto}B^{2(\alpha+\beta)+D}$. Eq.~(\ref{T1estimate}) suggests a monotonic behavior of $T_1^{-1}(B)$. For large enough magnetic fields, non-monotonic behavior may occur due to (possibly avoided) crossings of orbital levels, destructive interference of different mechanisms, or Van Hove singularities in the phonon density of states, see e.g.~Fig.~\ref{pic4} [\onlinecite{Droth2013,Hachiya2014,Bloch1957,Redfield1957,Bulaev2008}].

\begin{table}[t]
\centering
\caption{Specific examples for Eq.~(\ref{T1estimate}).\\
$\dag$ In (armchair) graphene nanoribbons, $\mathcal{H}_{\rm DEF}$ and $\mathcal{H}_{\rm BLC}$ correspond to the same parameters and $\mathcal{H}_{\rm DSP}$ does not give a net contribution, see Subsec.~\ref{T1GNR} or Ref.~[\onlinecite{Droth2013}].}
\begin{tabular}{llllll}
\hline
System&$\alpha$&$\beta$&$D$&$\eta$&$T_1^{-1}{\propto}B^{...}$\\
\hline
Graphene nanoribbon$^{\dag}$&1&1&1&1&5$\qquad$[\onlinecite{Droth2013}]\\
\hline
Bulk graphene ($\mathcal{H}_{\rm DEF}$)&1&0&2&1&4$\qquad$[\onlinecite{Struck2010}]\\
\hline
Bulk graphene ($\mathcal{H}_{\rm BLC}$)&0&0&2&1&2$\qquad$[\onlinecite{Struck2010}]\\
\hline
Bulk graphene ($\mathcal{H}_{\rm DSP}$)&0&0&2&2&0$\qquad$[\onlinecite{Struck2010}]\\
\hline
Bulk graphene ($\mathcal{H}_{\rm DSP}$)&0&0&2&1&2$\qquad$[\onlinecite{Struck2010}]\\
\hline
GaAs&0&1&3&1&5$\qquad$[\onlinecite{Khaetskii2001,Amasha2008}]\\
\hline
Silicon metal-oxide&1&1&3&1&7$\qquad$[\onlinecite{Xiao2010}]\\
\hline
\end{tabular}
\label{t3}
\end{table}

In graphene, we consider two mechanisms for the electron-phonon coupling, namely the deformation potential $\mathcal{H}_{\rm DEF}{=}g_1(u_{xx}{+}u_{yy})$ and the bond-length change $\mathcal{H}_{\rm BLC}{=}\tau\sigma_x(\tau A_x){+}\sigma_y(\tau A_y)$,
\begin{equation}
\mathcal{H}_{\rm EPC}{=}\mathcal{H}_{\rm DEF}{+}\mathcal{H}_{\rm BLC}\,.\label{EPC}
\end{equation}
In a minimal coupling picture, the form of $\mathcal{H}_{\rm BLC}$, where $(A_x,A_y){=}g_2(u_{xx}{-}u_{yy},{-}2u_{xy})$, highlights the equivalence of strain and a valley-dependent magnetic field, see Eq.~(\ref{Dirac}) and Refs.~[\onlinecite{Suzuura2002,Guinea2009}]. The strain tensor is given by $u_{ij}{=}(\partial_iu_j{+}\partial_ju_i)/2$ and the coupling constants are $g_1{\approx}30\,\text{eV}$ and $g_2{\approx}1.5\,\text{eV}$ [\onlinecite{Struck2010,Suzuura2002}]. We point out that all possible in-plane strains $u_{ij}$ ($i,j{=}1,2$) are considered by Eq.~(\ref{EPC}).

The admixture mechanism creates a lowest-order contribution for in-plane deformations $(u_x,u_y)$ but not for out-of-plane deformations $u_z$ since a local tilt of the graphene lattice does not give rise to finite $\mathcal{H}_{\rm EPC}$. Out-of-plane phonons can still induce a spin flip via the intrinsic spin-orbit interaction $\mathcal{H}_{\rm I}$, given in Eq.~(\ref{SOI}), or in higher order, where stretching effects lead to nonzero $\mathcal{H}_{\rm EPC}$. In the former case a small local tilt $(\partial_x,\partial_y)u_z$, where $|(\partial_x,\partial_y)u_z|{\ll}1$, changes the mutual orientation of pseudospin $\vec{\sigma}$ and real spin $\vec{s}$. As a consequence, the intrinsic spin-orbit coupling acquires spin-flip terms,
\begin{equation}
\tilde{\mathcal{H}}_{\rm I}{=}\lambda_{\rm I}\sigma_z\tau_z(s_z{-}\partial_xu_zs_x{-}\partial_yu_zs_y){=}\mathcal{H}_{\rm I}+\mathcal{H}_{\rm DSP}\,,\label{directSPC}
\end{equation}
and thus gives rise to direct spin-phonon coupling $\mathcal{H}_{\rm DSP}$ [\onlinecite{Rudner2010}]. The origin of this mechanism is illustrated in Fig.~\ref{pic2} (b). Below, we discuss the relaxation times of the electron spin in specific graphene QDs where the electron spectrum is gapped and where the valley degeneracy is lifted.

%

\subsection{Armchair graphene nanoribbons}\label{T1GNR}
\begin{figure}[t]\centering\includegraphics[width=1\linewidth]{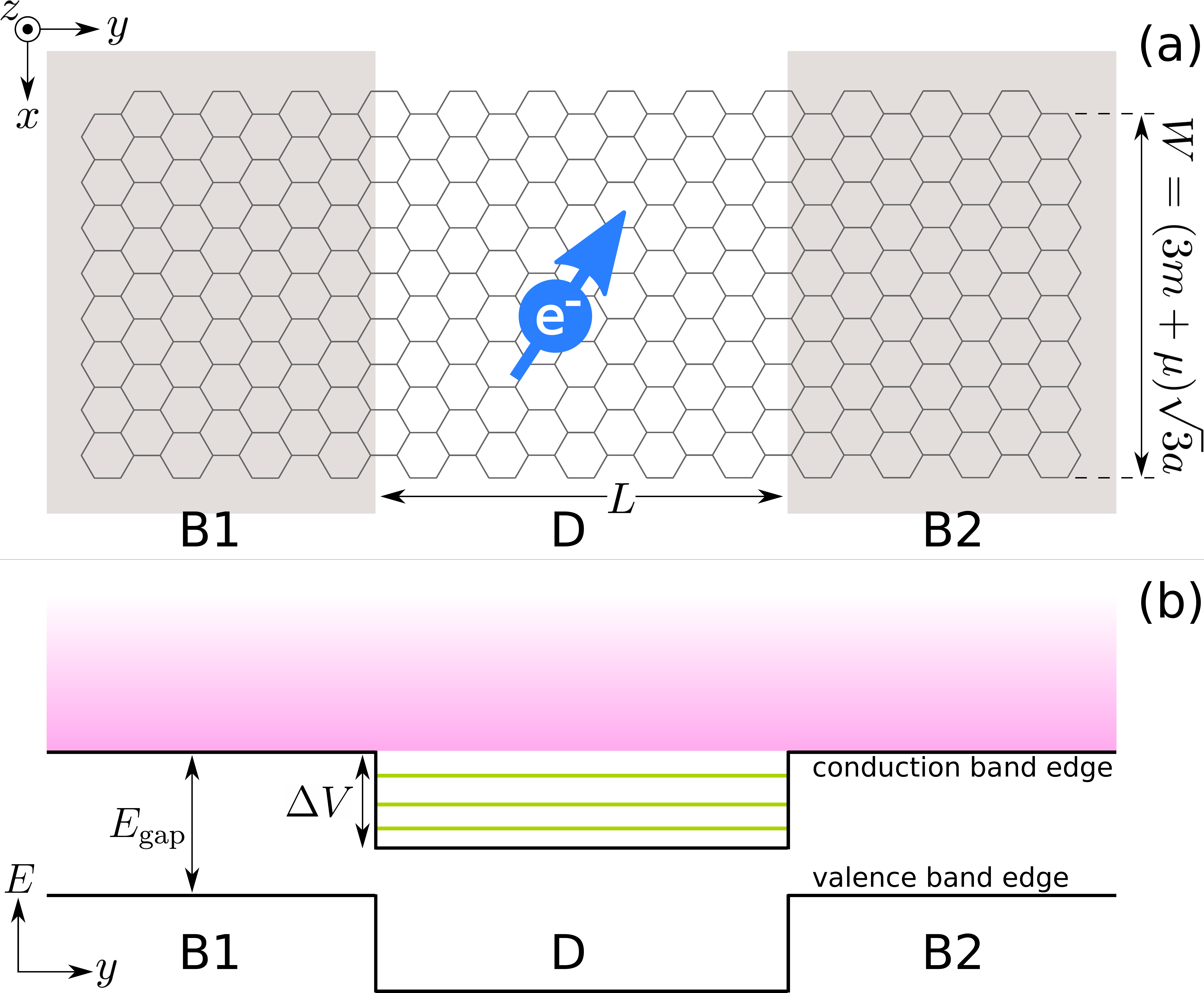}\caption{Sketch of (a) an armchair GNR and (b) its electronic spectrum. (a) Here, the width of the GNR is characterized by two integers, $m{=}3$ and $\mu{=}{-}1$. The resulting bandgap allows for electrostatic confinement along the $y$-axis. A potential $V(y)$ defines the barrier regions B1 and B2 (shaded) as well as the dot region D in between. (b) For $\Delta V{<}E_{\rm gap}$, there is one or more bound state(s) inside the quantum dot (green lines). A continuum of extended states exists for energies above the confinement potential (pink shade). Both bound and extended states provide virtual intermediate levels for the admixture mechanism responsible for spin relaxation.}\label{pic3}\end{figure}

In graphene nanoribbons (GNRs), electrons are naturally confined within the quasi one-dimensional structure and can move freely only along the longitudinal direction ($y$-axis in Fig.~\ref{pic3}). 

While the longitudinal wavenumber $q_y$ may vary continuously, the transverse wavenumber $q_n$ is discrete and depends on the ribbon width $W$ as well as the specific boundary conditions. In GNRs with lateral armchair boundaries, all components of the wavefunction spinor must vanish on the outermost atoms, which implies
\begin{equation}
q_n{=}\pi(n{-}\mu/3)/W\,,\label{transqx}
\end{equation}
where $n{=}0,{\pm}1,{\pm}2,...$ labels the transverse excitations. The width of the armchair GNR, $W{=}(3m{+}\mu)\sqrt{3}a$, is determined by $m{=}1,2,3,...$ and $\mu{\in}\{-1,0,+1\}$, see Fig.~\ref{pic3} (a) and Ref.~[\onlinecite{Brey2006}]. For a ribbon with $\mu{=}{\pm}1$, all transverse wavenumbers $q_n$ will be finite, thus leading to a gapped phonon spectrum with $E_{\rm gap}{=}2\hbar v_{\rm F}|q_0|{=}2\pi\hbar v_{\rm F}/3W$. For $W{=}30\,\text{nm}$, this amounts to $E_{\rm gap}{\approx}40\,\text{meV}$. The gap allows for electrostatic confinement of electrons along the longitudinal direction by virtue of an electrostatic potential $V(y)$. Analytical solutions for bound states are known if the potential has the piecewise constant form $V(y){=}\Delta V(\theta({-}y)+\theta(y{-}L))$, where $L$ is the longitudinal dimension of the quantum dot, see Fig.~\ref{pic3} (a) and Ref.~[\onlinecite{Trauzettel2007}]. At least one bound state exists if the confinement potential does not exceed the bandgap, $\Delta V{<}E_{\rm gap}$. Depending on $\Delta V$, there may be several bound states and independent of $\Delta V$, there is always a continuum of extended states energetically above $\Delta V$. Since $E_{\rm gap}$ is typically of the order of $\Delta V$, we consider both bound and extended states for virtual transitions within the admixture mechanism.

Due to the boundary conditions, the electronic states become superpositions of $K$ and $K'$ with equal weight on both valleys. This has important consequences. First, the valley degeneracy is lifted, thus circumventing the necessity of a complicated exchange interaction. Nevertheless, the orbital system is symmetric under time reversal and the spin states in a certain orbit form a Kramers pair at zero magnetic field, which suppresses spin relaxation due to Van Vleck cancellation. Furthermore, the direct spin-phonon coupling $\mathcal{H}_{\rm DSP}$ does not allow for spin relaxation as the according matrix element is proportional to the valley index $\tau$ and hence vanishes after summing up the contributions from $K$ and $K'$. The spin may still relax via $\mathcal{H}_{\rm EPC}$ and the admixture mechanism. In Eq.~(\ref{Fermi}), phonon emission [absorption] is proportional to $m_{\alpha,\vec{q}}{+}1$ [$m_{\alpha,\vec{q}}$], where the phonon number $m_{\alpha,\vec{q}}$ is determined by the Bose-Einstein distribution. Here, we assume phonon vacuum, i.e., pure phonon emission.

\begin{figure}[t]\centering\includegraphics[width=1\linewidth]{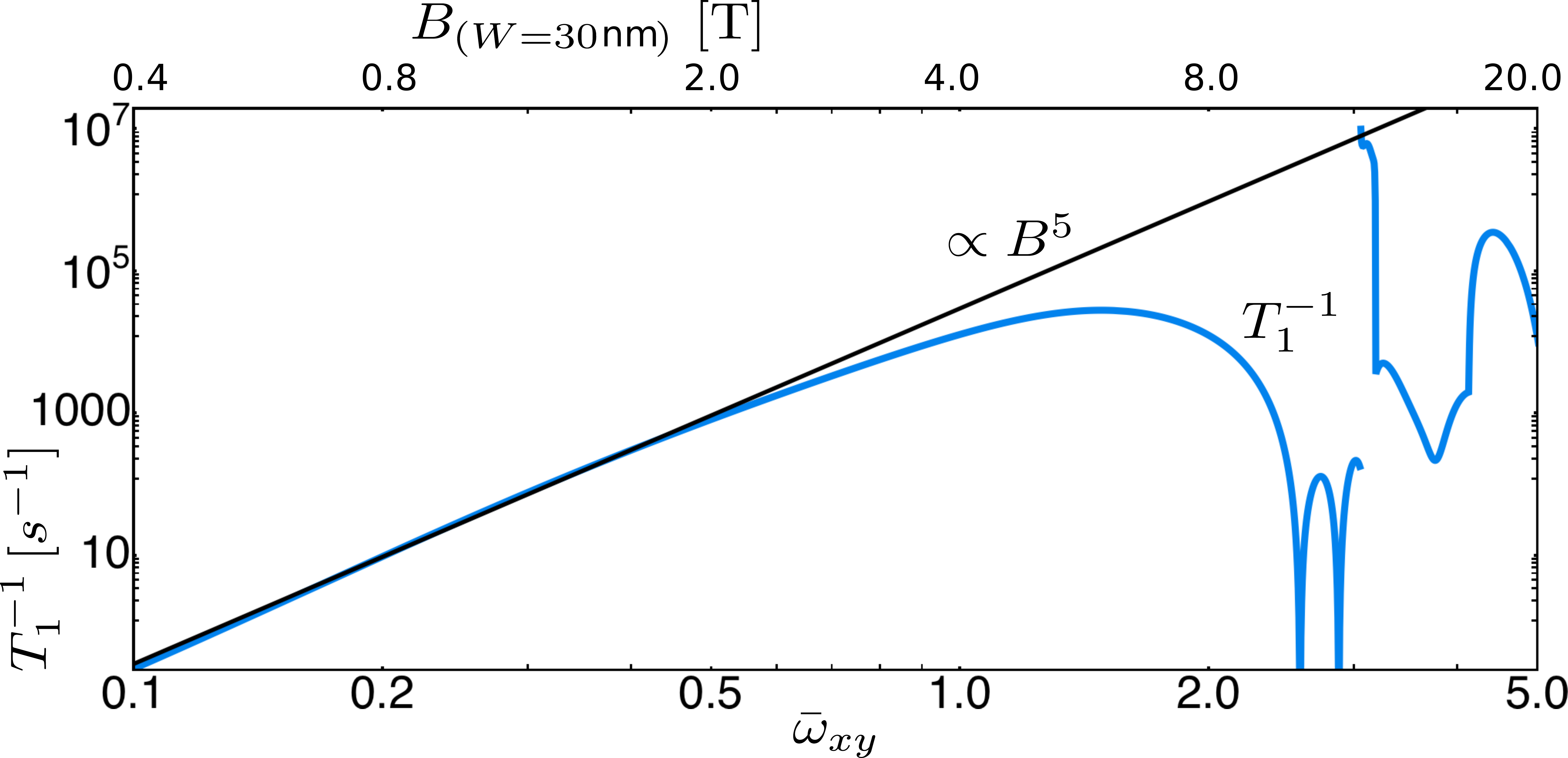}\caption{Spin relaxation rate $T_1^{-1}(B)$ of the QD ground state for $L/W{=}5$, $\Delta V{=}1.8\hbar v_{\rm F}q_0$, and free mechanical boundaries [\onlinecite{Droth2013}]. The doubly logarithmic scale highlights the behavior $T_1^{-1}{\propto}B^5$ for $\bar{\omega}{<}0.5$, in agreement with Eq.~(\ref{T1estimate}). The dimensionless phonon frequency $\bar{\omega}$ is explained in the main text. Destructive interference of processes originating from $\mathcal{H}_{\rm DEF}$ and $\mathcal{H}_{\rm BLC}$ causes two dips of $T_1^{-1}$ for $2{<}\bar{\omega}{<}3$ and a Van Hove singularity of the phonon density of states leads to a divergence around $\bar{\omega}{=}3$.}\label{pic4}\end{figure}

The spin relaxation depends on the confinement potential $\Delta V$, the QD aspect ration $L/W$, and the applied magnetic field $B$, which is assumed to be perpendicular to the graphene plane. Moreover, the mechanical boundary conditions determine the phonon spectrum and hence also the relaxation rate. In Fig.~\ref{pic4}, we plot the spin relaxation rate $T_1^{-1}(B)$ in blue as calculated with Eq.~(\ref{Fermi}) for $L/W{=}5$, $\Delta V{=}1.8\hbar v_{\rm F}q_0$, and free lateral boundaries [\onlinecite{Droth2011,Droth2013}]. In agreement with Eq.~(\ref{T1estimate}), the doubly logarithmic scale highlights the behavior $T_1^{-1}{\propto}B^5$ for $\bar{\omega}{<}0.5$, where $\bar{\omega}{=}\omega\sqrt{\rho_{\rm g}/Y}W$ is the dimensionless frequency of in-plane phonons with $Y$ and $\rho_{\rm g}$ as Young's modulus and the mass density of graphene, respectively [\onlinecite{Droth2011,Droth2013}]. Due to destructive interference of processes originating from $\mathcal{H}_{\rm DEF}$ and $\mathcal{H}_{\rm BLC}$, the relaxation rate features two dips for $2{<}\bar{\omega}{<}3$. A Van Hove singularity of the phonon density of states $\rho$ leads to a divergence around $\bar{\omega}{=}3$ --- a common effect for quasi one-dimensional systems [\onlinecite{Bulaev2008}].

Results for different values of $\Delta V$ and $L/W$, or for fixed mechanical boundary conditions can be found in Ref.~[\onlinecite{Droth2013}]. As one would expect, virtual transitions to extended states gain importance as $\Delta V/E_{\rm gap}$ decreases and even dominate over virtual transitions to bound states for $\Delta V/E_{\rm gap}{\to}0$. Fixed mechanical boundaries lead to a gapped phonon spectrum which excludes spin relaxation via single-phonon processes for $\bar{\omega}{<}2.06$ (or $B{<}8.25\,\text{T}$ for $W{=}30\,\text{nm}$).

As a quasi one-dimensional system, the armchair GNR is particularly interesting for spintronics-based quantum computing as it leaves two spatial dimensions for system periphery in a scaled-up device with multiple QDs along the GNR. The wave functions of bound states also extend to neighboring QDs and thus allow for multiple-qubit operations. Moreover, the small bandgap allows the coupling of non-neighboring QDs via detuning of the intermediate QDs, which allows for more efficient quantum computation since shuttling of qubits can be avoided and thus raises the threshold for fault-tolerant quantum computing [\onlinecite{Recher2010,Trauzettel2007,Svore2005,Braun2011}].

\subsection{Mono- and bilayer graphene disks}\label{mbGds}
Gate-tunable QDs with an electrostatic confinement potential can exist in 2D bulk mono- and bilayer graphene. Analytical solutions exist if the systems have circular symmetry in the graphene plane and the confinement potential has a rectangular profile, $V(r){=}\Delta V\theta(r-R)$, where $r$ is the radial coordinate and $R$ is the radius of the disk-shaped quantum dot. Due to the absence of intervalley scattering, the orbital degeneracy between valleys $K$ and $K'$ can be broken in a controlled way by the inclusion of a mass term --- a sublattice splitting for monolayer graphene or a layer bias in the case of bilayer graphene --- in combination with a magnetic field. In particular, the splitting of orbits in opposite valleys can be tuned with the applied magnetic field. In addition to a valley splitting, the mass term also gives rise to the required bandgap [\onlinecite{Recher2010,Recher2009}]. For magnetic fields, where the magnetic length $l_{\rm B}{=}\sqrt{\hbar/eB}$ is much smaller than $R$, the discrete QD spectrum converges to the bulk Landau levels. 
Below, we outline the analytical solution of the eigensystem both for mono- and for bilayer graphene. We first treat monolayer graphene disks and bilayer graphene afterwards.

As is obvious from Eq.~(\ref{spectrum}), a sublattice potential $\Delta$ will open up a bandgap of $E_{\rm gap}{=}2\Delta$. A sublattice potential arises when graphene lies on a substrate but the two sublattices couple differently to the substrate material as, e.g., for graphene on hBN, where $\Delta{>}100\,\text{meV}$ has been observed [\onlinecite{Hunt2013}]. While electron spin is neglected, here, the perpendicular component of the magnetic field is taken into account via minimal coupling to the motion of the charge carriers. Pseudospin can be included in the total angular momentum, $J_z{=}{-}i\partial_{\phi}{+}\sigma_z/2$, which commutes with the Hamiltonian. As a result of the circular symmetry, the azimuthal dependence can be separated away, leaving a differential equation for the radial dependence. Both for the dot region ($r{\leq}R$) and for the barrier region ($r{\geq}R$), the solutions are found to be hypergeometric functions. The matching condition for $r{=}R$ leads to a discrete spectrum of bound states. The valley degeneracy of $K$ and $K'$ is lifted for finite magnetic field and exceeds the Zeeman splitting for experimentally accessible parameters [\onlinecite{Recher2009}].

\begin{figure}[t]\centering\includegraphics[width=1\linewidth]{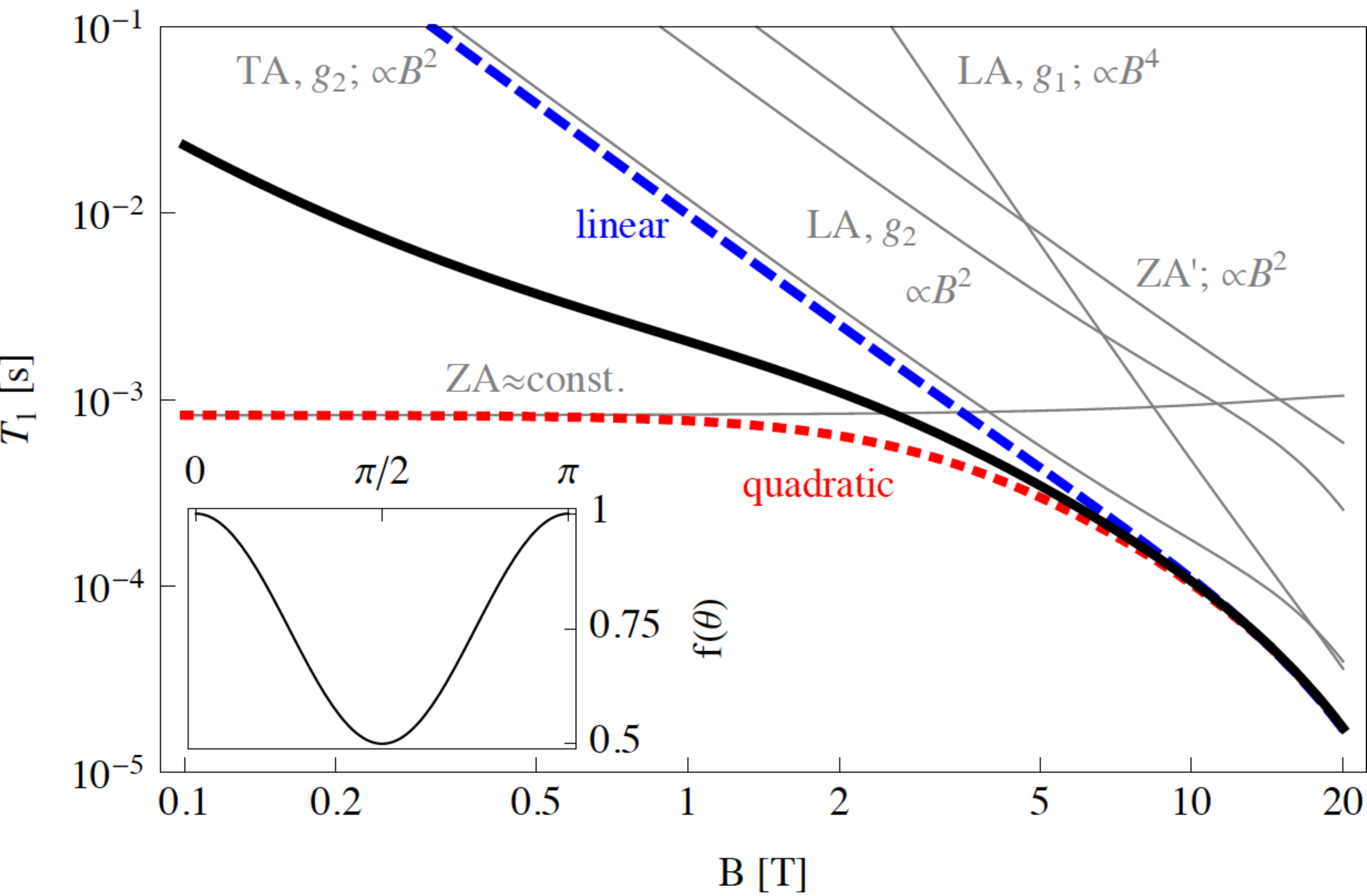}\caption{Spin relaxation times $T_1(B)$ for the ground state of a circular QD in monolayer graphene and a perpendicular magnetic field. The QD radius is $R{=}25\,\text{nm}$ and here, the bandgap $E_{\rm gap}{=}2\Delta$ is twice the confinement potential, $\Delta V{=}\Delta{=}260\,\text{meV}$. The coupling constants $g_1$ and $g_2$ indicate the associated mechanisms, $\mathcal{H}_{\rm DEF}$ and $\mathcal{H}_{\rm BLC}$, respectively. Both LA and TA in-plane modes disperse linearly while the out-of-plane mode has a transitions from quadratic (ZA) to linear (ZA') dispersion. The sum of all relaxation channels is represented by the red dotted, blue dashed, and black solid lines. A quadratic (linear) dispersion of the out-of-plane modes is assumed for the red (blue) curve, and a crossover between them for the black curve. The inset shows the universal dependence of $T_1^{-1}{\propto}f(\theta)$ on the inclination angle $\theta$ of the magnetic field [\onlinecite{Struck2010}]. \copyright APS.}\label{pic5}\end{figure}

As for spins in armchair GNR quantum dots, relaxation via the admixture mechanism and via direct spin-phonon coupling will be considered in the following. Assuming $\Delta{=}260\,\text{meV}$, no continuum of extended states has been considered in Ref.~[\onlinecite{Struck2010}]. As a result of the broken valley degeneracy, the spin states do not form a Kramers pair and Van Vleck cancellation does not occur in circular QDs in monolayer graphene. The spectrum of acoustic phonons consists of three branches, transverse (TA) and longitudinal (LA) in-plane modes, and one out-of-plane mode (ZA). Near the center of the Brillouin zone, the in-plane modes disperse linearly and the ZA modes quadratically with a crossover to linear behavior. Only the LA mode changes the size of the unit cell and is thus the only mode where the deformation potential ($\mathcal{H}_{\rm DEF}$) is nonzero. Both LA and TA deform the unit cell and hence generate a finite coupling via the bond-length change ($\mathcal{H}_{\rm BLC}$). ZA modes give rise to direct spin-phonon coupling ($\mathcal{H}_{\rm DSP}$). The according spin relaxation times for an electron in the QD ground state and phonon vacuum have been calculated in Ref.~[\onlinecite{Struck2010}] with specific parameters. The results are shown in Fig.~\ref{pic5}. For excited QD states, the relaxation times decrease quickly.

In a two-dimensional material, the minimal coupling prescription involves the perpendicular components of the $B$-field. But the Zeeman energy, which is transformed to lattice excitations by the mentioned couplings, depends on the entire magnetic field strength. As a result, all relaxation rates also depend on the inclination angle $\theta$ between the vector of the magnetic field and the normal vector of the graphene plane. For purely in-plane magnetic fields, the relaxation rates $T_1^{-1}$ are reduced by 50\% compared with the case of a perpendicular field, see inset in Fig.~\ref{pic5}.

In bilayer graphene, a gap opens if the electrostatic potential is different for the two layers. Then, an electrostatic confinement potential can be used to confine charge carriers in a quantum dot. Such systems are studied in experiment [\onlinecite{Allen2012,Goossens2012,Varlet2014,Tan2015}]. With appropriate top and bottom gates, the bandgap as well as the confinement potential can be tuned electrically and independently [\onlinecite{Recher2009}]. In a simple description, the system consists of two individual graphene layers, where one sublattice of the upper layer couples to one sublattice of the lower layer with an interlayer hopping amplitude $t_{12}{\approx}0.4\,\text{eV}$ [\onlinecite{Slawinska2010}]. The inversion symmetry of the bilayer system is broken by the potential difference between the distinct layers.

As for the case of monolayer graphene above, the problem can be solved analytically for a potential of the form $V(r){=}\Delta V\theta(r-R)$. Due to the contribution of the pseudospin of both layers, the total angular momentum is an integer here. The analytic solution of the eigensystem follows the same procedure as for the monolayer case but is more complicated because of the two coupled layers. The radial parts of the wavefunctions again turn out to be hypergeometric functions both inside and outside the dot region, and the matching condition at $r{=}R$ leads to a discrete spectrum of bound states. The valley degeneracy for $B{=}0$ can be lifted with a finite magnetic field perpendicular to the graphene plane and for $l_{\rm B}{\ll}R$, the energy spectrum approaches the Landau level structure. In particular, there are no Kramers pairs for a finite magnetic field perpendicular to the graphene plane.

To our knowledge, the spin relaxation time $T_1$ of a bound state inside the QD has not been estimated yet. For acoustic phonons neat the center of the Brillouin zone, the atoms within a unit cell do not vibrate against each other. Moreover, the interlayer Van der Waals bonds of bilayer graphene are much less rigid than the intralayer $\sigma$-bond. As a consequence, the change of the interlayer distance can be neglected and the individual layers can be treated completely independently for small wavenumbers. The deformation potential and the bond-length change within each layer are thus given by two copies of Eq.~(\ref{EPC}), one for each layer. Using the spin-orbit coupling of bilayer graphene, $T_1^{-1}$ can in principle be calculated via the admixture mechanism as for monolayer graphene [\onlinecite{Konschuh2012,Guinea2010}]. As for monolayer graphene, out-of-plane modes change the mutual orientation of real spin $\vec{s}$ and lattice related spins (pseudospin $\vec{\sigma}$ and for bilayer graphene also the layer spin $\vec{\mu}$ [\onlinecite{Konschuh2012}]). The resulting direct spin-phonon coupling might turn out to be more complicated than that for monolayer graphene. For coupling to the in-plane modes of bilayer graphene, Eqs.~(\ref{T1estimate}) and (\ref{EPC}) suggest a behavior $T_1^{-1}{\propto}B^2$ or ${\propto}B^4$, depending on the order of the dipole approximation for the specific mechanism.

\section{Magnetism in graphene nanoflakes}\label{GNF}
Both ferromagnetic and antiferromagnetic materials are important for data processing and storage (e.g. MRAM) [\onlinecite{Wolf2001}]. In contrast to typical magnetic elements like Fe, Ni, Co, and rare earths, carbon does not contain occupied d- or f-orbitals and pristine graphene is strongly diamagnetic [\onlinecite{Yazyev2010}]. Nevertheless, the prospects of magnetic ordering at room temperature and unmatched data density in monatomically thin layers motivate intense research efforts aimed at magnetism in graphene. Magnetic moments may appear in graphene because of vacancies [\onlinecite{Yazyev2007}], light or heavy adatoms [\onlinecite{Nair2012,McCreary2012,Zhang2012}], edge effects [\onlinecite{Tao2011,Son2006}], or molecular doping [\onlinecite{Hong2012,Nair2013}]. The resulting magnetic structures may be 0D, 1D, or 2D, depending on the origin of magnetism (e.g., isolated defects, GNR edges, or molecular doping of bulk graphene, respectively). The experimental detection of magnetic moments is possible via SQUID (superconducting quantum interference device) magnetometry [\onlinecite{Nair2012}], spin transport measurement [\onlinecite{McCreary2012}], or spin-sensitive STM (scanning tunneling microscopy) [\onlinecite{Hong2012,Decker2013}].

In the following, we focus on magnetic moments induced by vacancies or light adatoms. According to Lieb's theorem, the magnetic moment of the ground state of the Hubbard model for graphene is given by $\mu_{\rm B}|N_{\rm A}-N_{\rm B}|$, where $N_{\rm A}$ and $N_{\rm B}$ are the numbers of sublattice sites [\onlinecite{Lieb1989,FernandezRossier2007}]. Carbon atoms can be removed from the graphene lattice by means of irradiation with electrons or argon ions [\onlinecite{McCreary2012,Krasheninnikov2007,Robertson2013}]. Alternatively, light adatoms like H and F passivate the p$_z$ orbital of the carbon atom to which they bond, thus effectively removing this carbon atom from its sublattice. A hBN substrate may be used to stabilize the hydrogen adsorption on one sublattice and to suppress migration of the adatoms [\onlinecite{Hemmatiyan2014}]. Curie temperatures for the ferromagnetic ground state can exceed $300\,\text{K}$ [\onlinecite{Yazyev2007,Giesbers2013}].

\begin{figure}[t]\centering\includegraphics[width=1\linewidth]{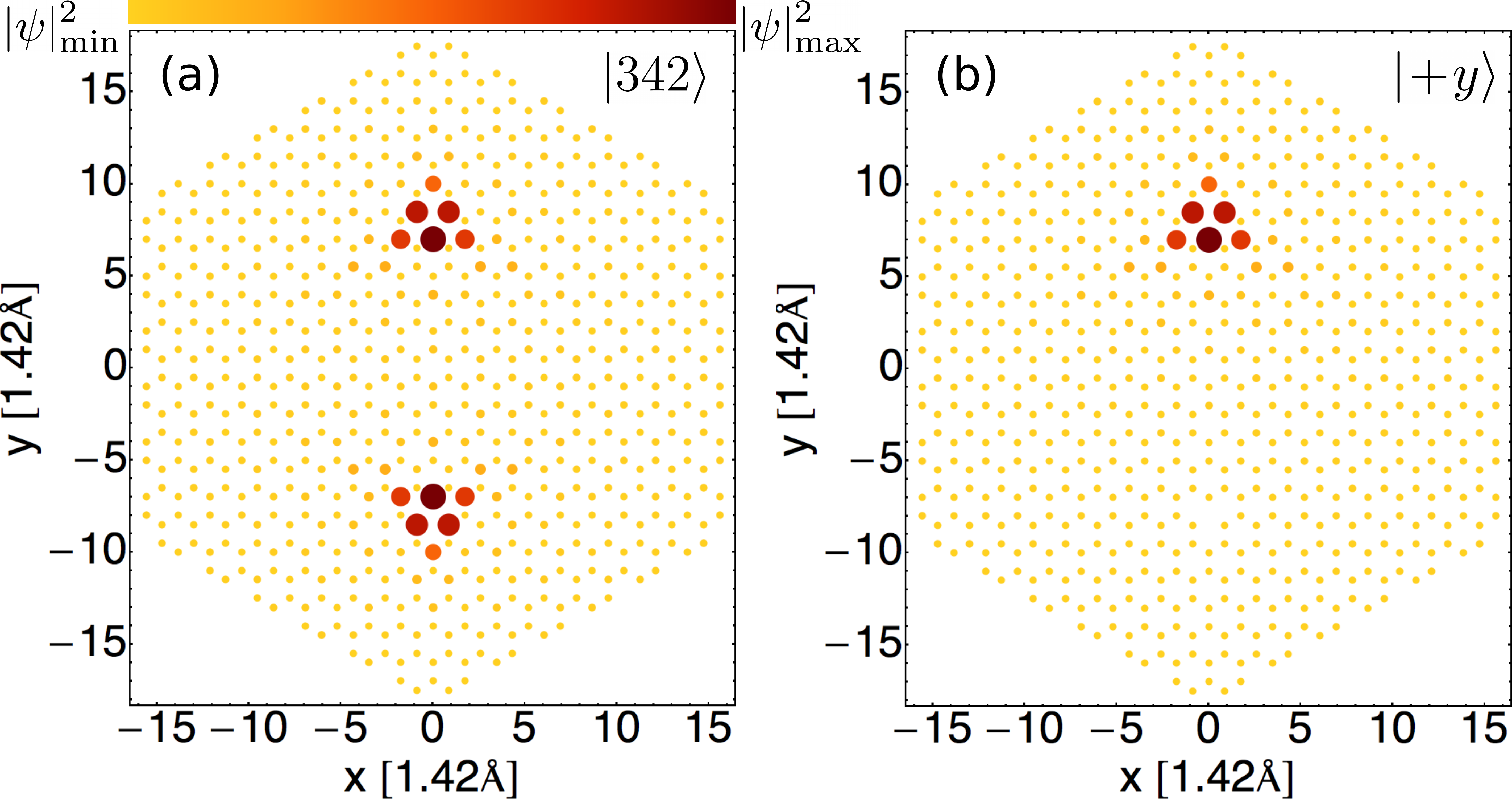}\caption{On-site probability densities for an armchair terminated hexagonal graphene nanoflake with a total 682 sites and two vacancies at $\vec{r}_{\rm vac}{=}(0,{\pm}8a)$. (a) Antibonding eigenstate $|1\rangle$. Here, we only consider nearest neighbor hopping such that the (anti-)bonding energy eigenstates $|341\rangle$ and $|342\rangle$ lie in the middle of the spectrum and have the same on-site probability densities. (b) localized vacancy state $|{+}y\rangle$. The probability density of $|{-}y\rangle$ looks similar yet mirrored about the $x$-axis [\onlinecite{Droth2015}].}\label{pic6}\end{figure}

The above defects, i.e., vacancies and light adatoms, are similar to quantum dots in that they lead to localized states. In Ref.~[\onlinecite{Droth2015}], we have studied the exchange coupling between two spin states localized at two separate yet nearby vacancies in a graphene nanoflake (GNF). The need for a bandgap does not play a role in GNFs since electrons are naturally confined within such quasi zero-dimensional structures. Specifically, we consider GNFs with hexagonal symmetry --- as found for flakes grown by chemical vapor deposition (CVD) [\onlinecite{Hamalainen2011,Phark2011}] --- and either zigzag or armchair terminations. Two atoms are removed at sites $\vec{r}_{\rm vac}{=}(0,{\pm}y)$ w.r.t.~the flake center such that the lattice retains the mirror symmetries $\mathcal{M}_x{:}x{\mapsto}{-}x$ and $\mathcal{M}_y{:}y{\mapsto}{-}y$, see Fig.~\ref{pic6}.

 We describe the electronic system with a full tight-binding calculation that involves hoppings $t_{ij}^{(n)}(B)$ between lattice sites $i$ and $j$ up to third nearest neighbors ($n{=}1,2,3$). A perpendicular magnetic field is included in the hoppings via the Peierls phase. The resulting spectrum is discrete and may or may not feature degeneracies, depending on the exact GNF configuration, $B$-field, and maximal order of nearest neighbor hoppings.

The two nearby vacancies can be interpreted as a double quantum dot (DQD), whose localized states form a bonding and an antibonding eigenstate within the GNF spectrum. If the energy splitting $\Delta$ between the bonding and antibonding eigenstates is much smaller than their energy differences to the remaining spectrum, the orbital part of the DQD can be described by
\begin{equation}
\mathcal{H}_{\rm DQD}{=}\begin{pmatrix}\bar{E}&t\\t^*&\bar{E}\end{pmatrix}\,,
\end{equation}
where we employ the localized states $\{|{+}y\rangle,\,|{-}y\rangle\}$ as a basis, $\bar{E}$ is the mean energy of the (anti-)bonding states, and where the intervacancy hopping with $|t|{=}\Delta{/}2$ may acquire a Peierls phase due to the magnetic field. Symmetry allows us to find the explicit form of $|{\pm}y\rangle$, see Fig.~\ref{pic6}. We complement this picture with an on-site Coulomb repulsion $U$ and aim to calculate the exchange coupling $J$, which energetically splits the singlet state $|S\rangle{=}\frac{1}{\sqrt{2}}(c^{\dagger}_{{+}y\uparrow}c^{\dagger}_{{-}y\downarrow}{-}c^{\dagger}_{{+}y\downarrow}c^{\dagger}_{{-}y\uparrow})|0\rangle$ from the triplet state $|T_0\rangle{=}\frac{1}{\sqrt{2}}(c^{\dagger}_{{+}y\uparrow}c^{\dagger}_{{-}y\downarrow}{+}c^{\dagger}_{{+}y\downarrow}c^{\dagger}_{{-}y\uparrow})|0\rangle$ and determines their mutual dynamics. Within this model and for $|t|{\ll}U$, we find that the singlet state is preferred, i.e., antiferromagnetic ordering, and
\begin{equation}
J{=}\frac{4|t|^2}{U}
\end{equation}
for the exchange coupling. Both $|t|$ and $U$, and hence $J$, depend on the perpendicular magnetic field that is included from the beginning. Typically, the magnetic field has a significant effect when the total magnetic flux through the GNF, $\Phi{=}AB$, where $A$ is the surface area of the nanoflake, reaches one flux quantum, $\Phi_0{=}h/2e$, where $h$ is Planck's constant and $e$ is the unit charge.

\begin{figure}[t]\centering\includegraphics[width=1\linewidth]{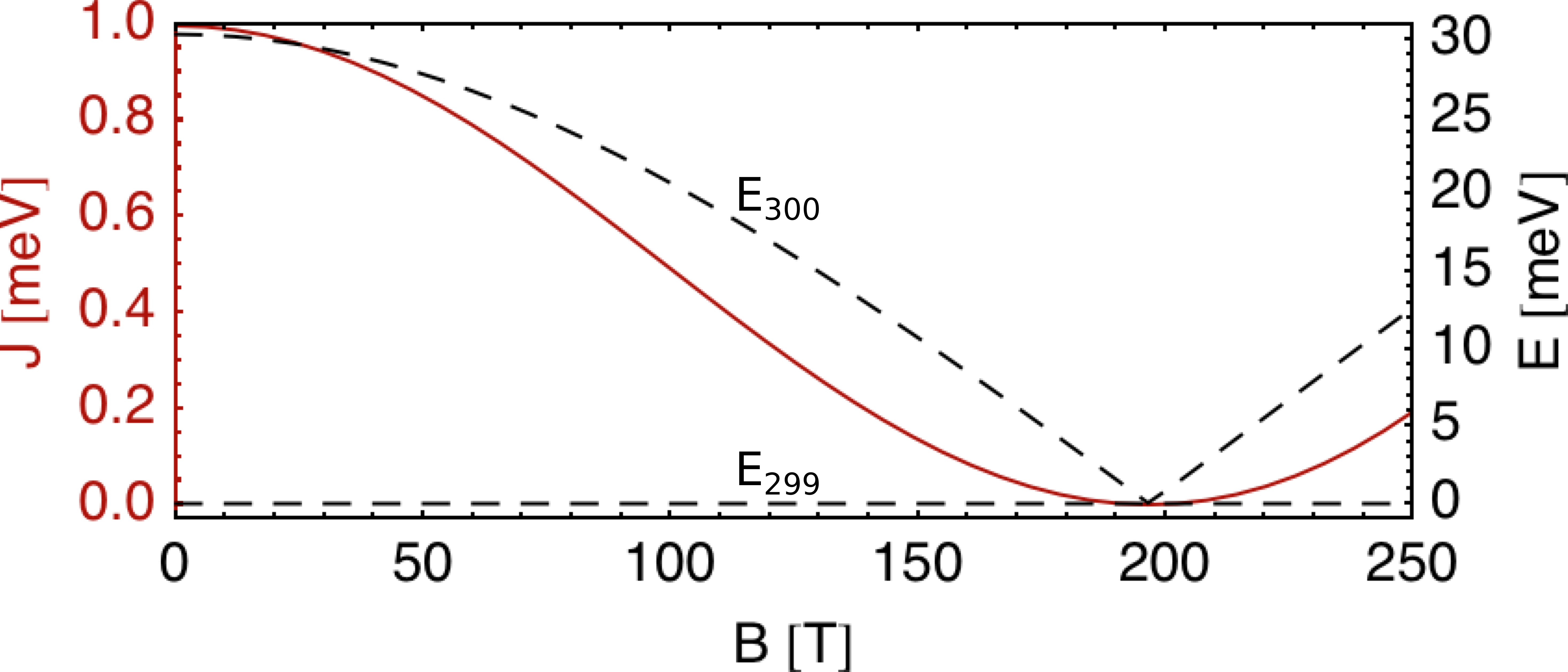}\caption{The exchange coupling $J$ (solid red line and left axis) and the eigenenergies (dashed black lines and right axis) of the according (anti-)bonding states are plotted against the perpendicular magnetic field. The shown data belongs to a nanoflake with zigzag terminations and a total 598 atoms [\onlinecite{Droth2015}]. Two vacancies are located at $\vec{r}_{\rm vac}{=}(0,{\pm}10a)$. Only nearest neighbor hopping is taken into account, such that the (anti-)bonding energy eigenstates $|299\rangle$ and $|300\rangle$ lie in the middle of the spectrum. One flux quantum passes through the flake at $B{=}145.8\,\text{T}$.}\label{pic7}\end{figure}

Depending on the overall size of the GNF, the distance between the vacancies, and the GNF termination (zigzag or armchair), the $B$-field dependence can be used to tune $J$ over several orders of magnitude. For some flake configurations, it is possible to tune the spectrum into a degeneracy, $\Delta(B){=}t(B){=}J(B){=}0$, thus switching off the antiferromagnetic order, see Fig.~\ref{pic7}. Such an \emph{in-situ} tunability of the exchange coupling is most promising for spintronics applications as it allows to change the magnetic behavior without preparing a new device. For flakes with armchair edges, we find that $J$ increases for GNFs of larger overall size and for smaller separation distances of the two vacancies. Bulk graphene can be viewed as the limit where the GNF diameter is much greater than the vacancy separation and where edge effects become negligible. We expect that the former trend saturates in this limit. The latter trend applies separately for two cases: (i) when the atom that has been removed for the vacancy at $(0,{+}y)$ belongs the A sublattice and (ii) when it belongs to the B sublattice. Such trends are not obvious for zigzag terminated flakes, where edge effects play a significant role for GNFs with up to ${\approx}10^4$ atoms [\onlinecite{Droth2015}]. N\'{e}el temperatures, $T_{\rm N}{\cong}J/k_{\rm B}$ with the Boltzmann constant $k_{\rm B}$, range from below $4\,\text{K}$ to beyond $300\,\text{K}$ and thus allow for experimental verification of these results, possibly via the techniques mentioned at the beginning of this section. Including a non-local Coulomb interaction might lead to ferromagnetic ordering, $J{<}0$, and possibly enable an \emph{in-situ} tunable crossover from antiferromagnetism to ferromagnetism [\onlinecite{Burkard1999}].

Our method relies on symmetry and hence the removal of an $A$ and a $B$ atom. Nevertheless, ferromagnetic ordering is expected if the two atoms are removed from the same sublattice and only on-site Coulomb interaction is considered [\onlinecite{Yazyev2010,Yazyev2008}]. We also point out that the exchange coupling $J$ can be used for coherent two-qubit operations, as required for universal quantum computing [\onlinecite{Petta2005,Loss1998,Shulman2012,Burkard1999}].

\section{Nuclear spins and graphene}\label{NSpins}
Naturally, only 1\% of all carbon atoms belong to the isotope $^{13}$C, which is the only stable carbon isotope with finite nuclear spin, see Table \ref{t1}. This abundance can be lowered by isotopic purification [\onlinecite{Balasubramanian2009}]. Possibly, this is the reason why the combination of nuclear spins and graphene has so far received little attention from the scientific community. However, there are at least three situations where nuclear spins become important for spintronics in graphene. (i) First, it has been found theoretically that $T_1$ relaxation times of electron spins in graphene QDs can exceed $100\,\text{$\mu$s}$ for certain parameters and if nuclear spins are neglected, see Figs.~\ref{pic4} and \ref{pic5} or Refs.~[\onlinecite{Droth2013,Struck2010}]. In this case, the spin dephasing time $T_2$ might be limited by the coupling to nuclear spins of the $^{13}$C nuclei. (ii) Then, state-of-the-art graphene QDs often employ a graphene (bi-)layer encapsulated by bottom and top sheets of hexagonal boron nitride. Neither boron nor nitrogen posses any stable isotopes without nuclear spin, see Table \ref{t1}. The spin of an electron in such an encapsulated graphene QD is in immediate contact with the nuclear spins of hBN. (iii) Finally, the nuclear spin of an impurity, of an adatom, or of $^{13}$C itself might be interesting for spintronics. Natural monolayer graphene contains one nuclear spin per $(16\,\text{\AA})^2$ surface area. This density can be controlled by enriching or depleting $^{13}$C. Possibly, control and coupling of nuclear spins in the graphene sheet or on top of it can be achieved. Ultimately, this could lead to 2D nuclear spintronics with similarities to nitrogen vacancy centers in diamond, where $^{13}$C nuclear spins have been used for quantum operations [\onlinecite{Hilser2012,Bassett2014}].

The Hamiltonian for an ensemble of nuclear spins in a solid has five different contributions [\onlinecite{Coish2009}],
\begin{equation}
\mathcal{H}_{\rm nuc}{=}\mathcal{H}_{\rm z}+\mathcal{H}_{\rm hf}+\mathcal{H}_{\rm orb}+\mathcal{H}_{\rm dd}+\mathcal{H}_{\rm q}\,.
\end{equation}
The term $\mathcal{H}_{\rm z}$ arises due to the Zeeman energies of nuclear spins in a magnetic field $B$. The second term describes the hyperfine interaction between electron spins and nuclear spins. It is the sum of two terms, $\mathcal{H}_{\rm hf}{=}\mathcal{H}_{\rm c}{+}\mathcal{H}_{\rm a}$. The contact (isotropic) hyperfine interaction can be modeled as an Overhauser field, $B_{\rm N}$, that acts on the electron spin $\vec{s}$ in a similar way as an external magnetic field,
\begin{equation}
\mathcal{H}_{\rm c}{=}\left(\sum_kA_k\vec{I}_k\right)\cdot\vec{s}{=}g\mu_{\rm B}\vec{B}_{\rm N}\cdot\vec{s}\,.\label{Overhauser}
\end{equation}
Here, $\vec{I}_k$ is the spin operator for the nucleus at site $k$ and the coupling strength $A_k$ is proportional to the local probability density of the electron and the gyromagnetic ratio of the nucleus at site $k$ [\onlinecite{Hanson2007,Coish2009,Overhauser1953,Khaetskii2002}]. Eq.~(\ref{Overhauser}) vanishes for $\pi$-band electrons (in particular those of graphene) as their probability densities at the nuclear sites are zero due to odd symmetry. In this case (in particular for graphene), the anisotropic hyperfine interaction, $\mathcal{H}_{\rm a}$, and the coupling of nuclear spin and electron orbit, $\mathcal{H}_{\rm orb}$, dominate electron-nuclear coupling. In addition to $\mathcal{H}_{\rm hf}$, also $\mathcal{H}_{\rm orb}$ can mediate a coupling of nuclear and electron spins if spin-orbit interaction is included. The nuclear dipole-dipole Hamiltonian, $\mathcal{H}_{\rm dd}$, directly couples the magnetic moments of distinct nuclear spins. And nuclei with spin $I{>}\frac{1}{2}$ (thus excluding $^{13}$C) interact with an electric field gradient via the nuclear quadrupolar coupling, $\mathcal{H}_{\rm q}$. All these terms are discussed in detail in Ref.~[\onlinecite{Coish2009}].

Spintronics with nuclear spins --- point (iii) in our list above --- would rely on the coupling of nuclear spins to external fields, which is achieved by $\mathcal{H}_{\rm z}$ and $\mathcal{H}_{\rm q}$, as well as the coupling between different nuclear spins, which is mediated by $\mathcal{H}_{\rm dd}$. To our knowledge, the effect of nuclear spins from hBN on the electron spin in encapsulated graphene QDs --- point (ii) --- have not been studied, yet. In the remainder of this section, we review studies on point (i), the decoherence of the electron spin due to hyperfine interaction with the nuclear spins of $^{13}$C.

The contact hyperfine interaction vanishes for flat graphene and indirect coupling of electron and nuclear spins via $\mathcal{H}_{\rm orb}$ and spin-orbit interaction is expected to be small and hence neglected. Consequently, the coupling between electron and nuclear spins in graphene is given by
\begin{equation}
\mathcal{H}_{\rm hf}{=}\mathcal{H}_{\rm a}{=}\sum_kA_{k,x}I_{k,x}s_x{+}A_{k,y}I_{k,y}s_y{+}A_{k,z}I_{k,z}s_z\,,\label{HF}
\end{equation}
where $k$ indexes the sites of $^{13}$C isotopes. The coupling constants $A_{k,i}$ are proportional to the probability density of the electron at site $k$ and the respective coupling strength $A_i$, with $A_x{=}A_y{=}{-}\frac{A_z}{2}{\approx}{-}0.3\,\text{$\mu$eV}$ [\onlinecite{Yazyev2008,Fischer2009}]. The $z$-axis is chosen to be perpendicular to the graphene sheet. In contrast, the coupling strength of the (contact) hyperfine interaction in GaAs ranges from $74\,\text{$\mu$eV}$ to $96\,\text{$\mu$eV}$ (depending on the specific isotope), thus exceeding that of graphene by more than one order of magnitude [\onlinecite{Coish2009}]. The importance of the hyperfine interaction in graphene has been studied by comparing CVD-grown samples consisting entirely of $^{12}$C and $^{13}$C, respectively [\onlinecite{Wojtaszek2014}]. The experimental results indicate that in contrast to the cases of GaAs or carbon nanotubes, hyperfine interactions with $^{13}$C nuclear spins have a negligible effect on spin transport relaxation times in CVD-grown graphene [\onlinecite{Coish2009,Churchill2009(2),Churchill2009}].

\begin{figure}[t]\centering\includegraphics[width=1\linewidth]{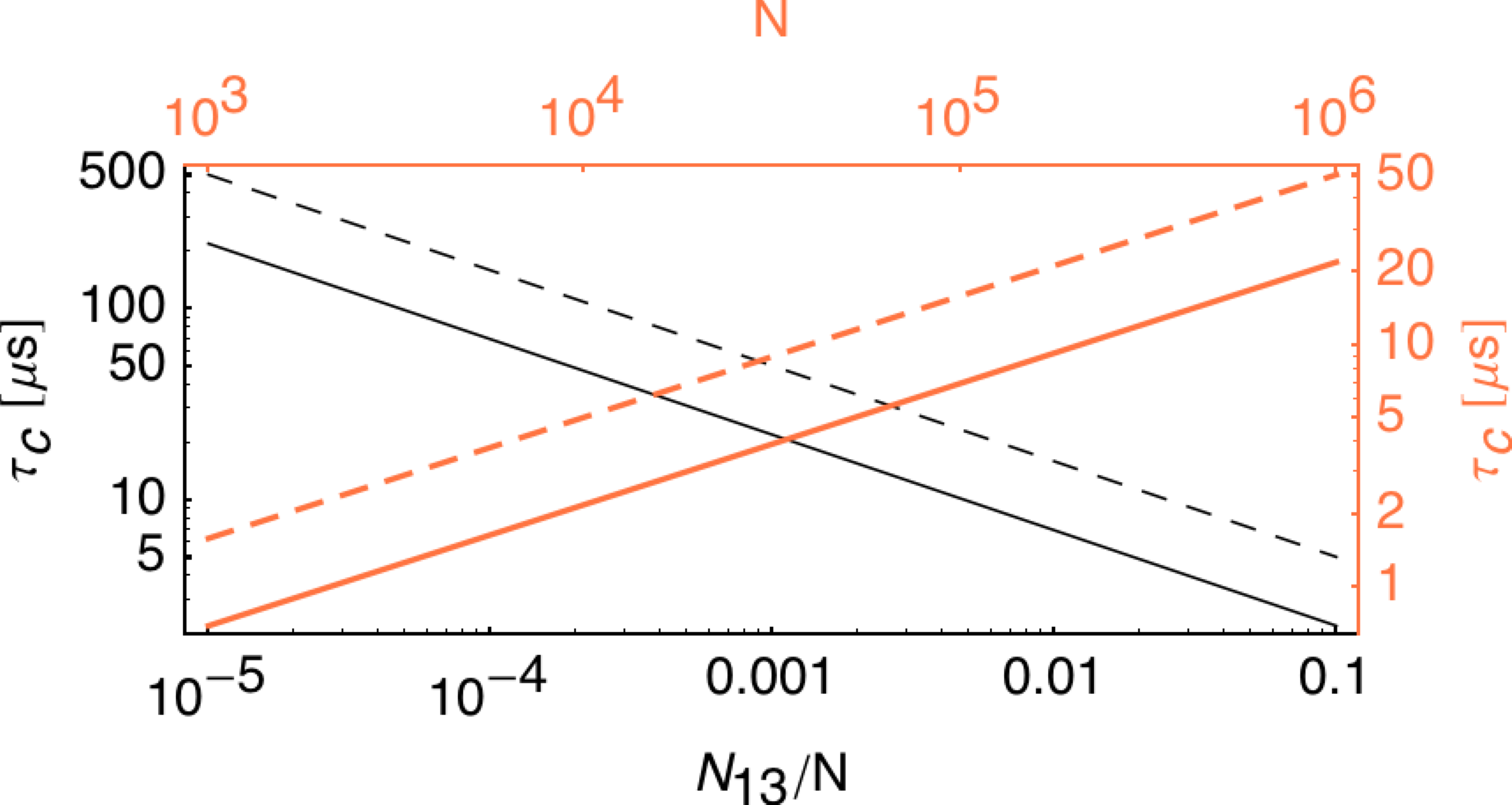}\caption{We plot Eq.~(\ref{tauc}) for various parameters on a doubly logarithmic scale. The longitudinal hyperfine coupling strength in graphene is $A_z{\approx}0.6\,\text{$\mu$eV}$ [\onlinecite{Fischer2009}]. The polarization of the nuclear spin system is assumed to be $p{=}0$ for the solid lines and $p{=}0.9$ for the dashed ones. For the thin black lines (left and bottom axes), we vary the relative $^{13}$C abundance but fix the total number of atoms to $N{=}10^5$, which corresponds to a square QD with an approximate size of ${\approx}(50\,\text{nm})^2$. For the thick orange lines (right and top axes), we vary $N$ and fix the relative $^{13}$C abundance to its natural value, $\frac{N_{13}}{N}{=}0.01$.}\label{pic8}\end{figure}

The interaction of an electron spin in a graphene QD with a certain number of nuclear spins, $N_{13}$, will affect the time evolution of its transversal component $\langle s_+\rangle(t)$, where $s_+{=}s_x{+}is_y$. If $N_{13}{\gg}1$, the nuclear spins form a bath and one can use the central limit theorem to obtain $\langle s_+\rangle(t){=}\langle s_+\rangle(0)e^{-\frac{t^2}{\tau_{\rm c}^2}}$ [\onlinecite{Fischer2009}]. In the presence of a perpendicular magnetic field $B_z{\gg}2.6\,\text{mT}$, the electron Zeeman energy is much greater than the transversal hyperfine coupling strength $|A_{x,y}|{\approx}0.3\,\text{$\mu$eV}$. Then, the characteristic decoherence time $\tau_{\rm c}$ is given by
\begin{equation}
\tau_{\rm c}{=}\frac{2\hbar}{\sqrt{1-p^2}}\frac{N}{\sqrt{N_{13}}A_z}\,,\label{tauc}
\end{equation}
where $0{\leq}p{\leq}1$ denotes the polarization of the nuclear spin system and $N$ is the total number of carbon atoms, i.e., $\frac{N_{13}}{N}{=}0.01$ for natural carbon [\onlinecite{Fischer2009}]. Typical values for $\tau_{\rm c}$ exceed $1\,\text{$\mu$s}$, see Fig.~\ref{pic8}. In a further study, it was found that in the same situation, the longitudinal spin component $\langle s_z\rangle(t)$ is conserved up to small corrections, which oscillate with a frequency determined by the hyperfine interaction [\onlinecite{Fuchs2012}]. If less than roughly 10 nuclear spins interact with the electron spin in the graphene QD, the nuclear spin system cannot be modeled as bath but allows for an exact treatment. Then, decoherence times depend on the exact position and orientation of the nuclear spins, and typically lie in the regime of $1\,\text{ms}$ [\onlinecite{Fuchs2013}].

\section{Experimental progress}\label{Exp}
Spintronics with individual electron spins in semiconductors has evolved dramatically from basic concepts to universal quantum control of spin states and quantum registers with up to four QDs [\onlinecite{Wolf2001,Awschalom,Awschalom2013,Jansen2012,Hanson2007,Petta2005,Loss1998,Shulman2012,Takakura2014}]. Isotopic purification enables coherence times beyond $1\,\text{s}$ at low temperature in silicon and beyond $1\,\text{ms}$ at room temperature in diamond [\onlinecite{Tyryshkin2012,Balasubramanian2009}].

Despite the challenge of efficient spin injection into graphene through tunneling barriers, remarkable improvements in spin transport have been achieved over the past few years [\onlinecite{Han2014,Guimaraes2012,Tombros2007}]. In contrast to electron mobility, the spin lifetime seems to be rather insensitive to charged impurity scattering [\onlinecite{Han2012}] and can exceed $1\,\text{ns}$ at room temperature [\onlinecite{Drogeler2014}], even for CVD-grown graphene [\onlinecite{Kamalakar2015}]. The small intrinsic spin-orbit interaction makes it practically impossible to observe the spin Hall effect (SHE) in graphene [\onlinecite{Kane2005}]. But controlled hydrogenation can increase the spin-orbit interaction by three orders of magnitude and thus enable the observation of a SHE at room temperature and zero magnetic field [\onlinecite{Balakrishnan2013}]. Last year, a quantum spin Hall state has been observed in graphene with a large, tilted magnetic field [\onlinecite{Young2014}].

Graphene QDs have been studied thoroughly w.r.t.~their orbital spectrum [\onlinecite{Liu2010,Allen2012,Goossens2012,Ponomarenko2008}], spin-filling sequence [\onlinecite{Guttinger2010}], and charge relaxation times [\onlinecite{Volk2013}]. Moreover, high-quality bilayer QDs, encapsulated in hBN and equipped with electric gates, allow for electrostatic confinement as discussed in Subsec.~\ref{mbGds}. Small QDs in the $1\,\text{nm}$ range can be created by electroburning and exhibit Coulomb blockade at room temperature due to addition energies as large as $1.6\,\text{eV}$ [\onlinecite{Barreiro2012}]. Slightly larger graphene nanoflakes, epitaxially grown on Ir(111) and intercalated with oxygen, are reported to exhibit a linear spectrum [\onlinecite{Jolie2014}]. Despite these promising proceedings with graphene-based QDs, Pauli blockade --- a basic step towards spintronics [\onlinecite{Ciorga2000}] and already achieved in carbon nanotubes [\onlinecite{Churchill2009(2),Churchill2009}] --- has (to our knowledge) not been observed in graphene, yet.

Gapped armchair GNRs, Subsec.~\ref{T1GNR}, and symmetrical graphene nanoflakes, Sec.~\ref{GNF}, rely on precise edges. Atomically accurate armchair GNRs can be produced bottom-up, with widths up to $4\,\text{nm}$, from aromatic precursor molecules [\onlinecite{Cai2010,Huang2012}] or top-down. One top-down technique that leads to precise edges is the unzipping of carbon nanotubes with an abruptly expanding nitrogen gas inside the nanotubes [\onlinecite{MorelosGomez2012}]. Another possibility is the electron-beam induced mechanical rupture of bulk graphene [\onlinecite{Kim2013}]. The latter method can be used to produce armchair edges as well as zigzag edges and can be applied in high vacuum, thus leading to minimally contaminated samples. Beyond this, clean structuring methods for graphene include nano-etching of suspended graphene [\onlinecite{Sommer2015}] and the use of a silicon atom as a monatomic chisel [\onlinecite{Wang2014(2)}].


\section{Conclusion and Outlook}\label{CandO}
We have given an overview of spintronics with electron spin states in graphene quantum dots and nanoflakes. Long coherence times can be expected because of weak sources of decoherence, i.e., small coupling strengths of the spin-orbit and hyperfine interactions, and a marginal abundance of nuclear spins in carbon. We have explained spin relaxation via spin-orbit mediated coupling to acoustic phonons and we have introduced a generic formula for the spin relaxation time $T_1$ for graphene and other systems. Valley degeneracy and the lack of a bandgap in bulk graphene pose challenges to spin operations in graphene quantum dots. Specific systems where these challenges can be overcome have been reviewed and, if applicable, studies on their spin relaxation times have been discussed. Magnetism can be induced in graphene by several means. Here, we have focused on magnetism in graphene nanoflakes and \emph{in-situ} tunability of antiferromagnetism in symmetric nanoflakes with defects. We have summarized studies on the interaction of electron and nuclear spins and finally, we have pointed out recent experimental achievements relevant for spintronics with graphene quantum dots. 

We believe that graphene has the potential for a top-notch spintronics material that can outperform established materials like silicon or GaAs for specific applications. The observation of the Pauli blockade represents a fundamental step towards spintronics with graphene quantum dots. In principle, the necessary architectures exist already but suffer from poor tunability of the source and drain barriers and inefficient spin injection. Similarly, the components required for electrostatic confinement of electrons in armchair GNR quantum dots are already available but have not been integrated into one device, yet. With view to the rapid progress of device fabrication, we anticipate significant progress in these fields within the next 2-3 years. Ultimately, we see spintronics in graphene as a building block for 2D-heterostructured spintronics devices composed of (functionalized) graphene and other 2D materials like TMDCs and hBN.


\section{Acknowledgements}
We thank the European Science Foundation and the Deutsche Forschungsgemeinschaft (DFG) for support within the EuroGRAPHENE project CONGRAN and the DFG for funding within SFB 767 and FOR 912.


\begin{thebibliography}{}
\bibitem{footnote}
We use upper case indices for graphene Hamiltonians and lower case indices for general Hamiltonians.
\end{thebibliography}
\end{document}